\documentclass[journal=ancac3,manuscript=article]{achemso}
\usepackage[version=3]{mhchem} % Formula subscripts using \ce{}
\usepackage{textcomp, gensymb}
\usepackage{xcolor}

\usepackage{float} 

\usepackage{caption}
\usepackage{subcaption}

\author{Elisa Ballin}
\affiliation{Dipartimento di Fisica, Sapienza Università di Roma, Piazzale A. Moro 5, 00185 Roma, Italy}
\alsoaffiliation{CNR-ISC, Uos Sapienza, Piazzale A. Moro 5, 00185 Roma, Italy}
\altaffiliation{These authors contributed equally to this work}
\email{elisa.ballin@uniroma1.it}
\author{Francesco Brasili}
\affiliation{CNR-ISC, Uos Sapienza, Piazzale A. Moro 5, 00185 Roma, Italy}
\alsoaffiliation{Dipartimento di Fisica, Sapienza Università di Roma, Piazzale A. Moro 5, 00185 Roma, Italy}
\altaffiliation{These authors contributed equally to this work}
\author{Tommaso Papetti}
\affiliation{Dipartimento di Fisica, Sapienza Università di Roma, Piazzale A. Moro 5, 00185 Roma, Italy}
\alsoaffiliation{CNR-ISC, Uos Sapienza, Piazzale A. Moro 5, 00185 Roma, Italy}
\author{Jacopo Vialetto}
\affiliation{Dipartimento di Chimica "Ugo Schiff", Università di Firenze,  Sesto Fiorentino (FI), 50019 Italy}
\alsoaffiliation{Consorzio per lo Sviluppo dei Sistemi a Grande Interfase (CSGI), via della Lastruccia 3, Sesto Fiorentino (FI), 50019, Italy}
\author{Michael Sztucki}
\affiliation{ESRF, The European Synchrotron, 71 avenue des Martyrs CS40220, 38043 Grenoble Cedex 9, France}
\author{Simona Sennato}
\affiliation{CNR-ISC, Uos Sapienza, Piazzale A. Moro 5, 00185 Roma, Italy}
\alsoaffiliation{Dipartimento di Fisica, Sapienza Università di Roma, Piazzale A. Moro 5, 00185 Roma, Italy}
\author{Marco Laurati}
\affiliation{Dipartimento di Chimica "Ugo Schiff", Università di Firenze,  Sesto Fiorentino (FI), 50019 Italy}
\alsoaffiliation{Consorzio per lo Sviluppo dei Sistemi a Grande Interfase (CSGI), via della Lastruccia 3, Sesto Fiorentino (FI), 50019, Italy}
\email{marco.laurati@unifi.it}
\author{Emanuela Zaccarelli}
\affiliation{CNR-ISC, Uos Sapienza, Piazzale A. Moro 5, 00185 Roma, Italy}
\alsoaffiliation{Dipartimento di Fisica, Sapienza Università di Roma, Piazzale A. Moro 5, 00185 Roma, Italy}
\email{emanuela.zaccarelli@cnr.it}

\title[]
  {Star-like thermoresponsive microgels: a new class of soft nanocolloids}

\abbreviations{DLS, SAXS, PNIPAM, BIS, EGDMA}

\begin{document}

%\begin{tocentry}

%\includegraphics[width=\textwidth]{toc-articolo.png}
%\end{tocentry}

\begin{abstract}
We provide experimental and numerical evidence of a new class of soft nanocolloids: star-like microgels with thermoresponsive character. This is achieved by using the standard precipitation polymerization synthesis of poly(N-isopropylacrylamide) (PNIPAM) microgels and replacing the usually employed crosslinking agent, N,N'-methylenebis(acrylamide) (BIS), with ethylene glycol dimethacrylate (EGDMA). The fast reactivity of EGDMA combined with its strong tendency to self-bind produces colloidal networks with a central, crosslinker-rich core, surrounded by a corona of long, crosslinker-free arms. 
These novel star-like microgels fully retain PNIPAM thermoresponsivity and undergo a volume phase transition at a temperature $\sim 32\,\celsius$ that is very sharp as compared to standard PNIPAM-BIS microgels, independently of crosslinker content. Dynamic light scattering and small angle X-ray scattering experiments are compared to extensive simulation results, based on ideal star polymers as well as on state-of-the-art monomer-resolved simulations, offering a microscopic evidence of the star-like internal structure of PNIPAM-EGDMA microgels. This can be described by a novel model for the form factors combining star and microgel features. The present work thus bridges the fields of star polymers and microgels, providing the former with the ability to respond to temperature via a facile synthetic route that can be routinely employed, opening the way to exploit these soft particles for a variety of fundamental studies and applicative purposes. 
%\begin{figure}[h!]
%\centering
%\includegraphics[scale=1]{TOC}
%\label{fig:toc}
%\end{figure}
\end{abstract}

\section{Introduction}
Star polymers are nanocolloids made of multiple arms radiating from a central core; this class of polymers has gathered in the past significant attention both for their fundamental theoretical interest~\cite{likos1998star,vlassopoulos2004colloidal} and for their potential applications in various fields, including biomedicine, environmental remediation, and advanced materials, because of the large spectrum of functionalisations available for the polymeric arms\cite{renStarPolymers2016,wuStarPolymersAdvances2015}. Star polymers are the prototype of ultrasoft particles, with an interesting logarithmic dependence of the interaction potential on interparticle distance~\cite{likos2006soft}, which arises in the limit of very small core with respect to the length of the arms. The synthesis of multi-arm stars is however quite involved~\cite{roovers1989synthesis} and requires deep chemical expertise; this difficulty, together with the fact that the polymers used are typically dispersed in organic solvents, has partially hindered the availability of the samples and their wide spreading as a model system for experimental work. Furthermore, while examples of pH-responsive multi-arm star polyelectrolytes are present\cite{connal2008}, only a few recent examples of thermoresponsive star-like systems have been reported~\cite{lang2017,Cao2018,okaya2020}. These are however limited to small arm numbers and by a synthetic process that retains a high level of complexity. Thereby, the possibility to easily synthesize star-like particles dispersed in water, and even to provide them thermoresponsiveness, would be a major breakthrough in the field.
On the other hand, microgels are soft colloidal particles consisting of a crosslinked polymer network, that are obtained by a much simpler synthesis. They are often dispersed in water and display responsive characteristics that are desirable for star polymers. Indeed, their polymeric nature endows them with the ability to respond to various external stimuli, such as light, pH, ionic strength, and temperature. The most widely investigated microgels are those based on poly(N-isopropylacrylamide) ({PNIPAM}), a thermoresponsive polymer that exhibits a coil-to-globule transition at about $32\celsius$ due to a change in solvent quality\cite{kubotaSinglechainTransitionPolyNisopropylacrylamide1990}. 
This enables PNIPAM-based  microgels to reversibly swell or shrink, modulating their effective volume fraction as a function of temperature, through the so-called Volume Phase Transition (VPT)~\cite{fernandez2011microgel}. 
PNIPAM-based microgels are typically synthesized via free radical precipitation polymerization~\cite{pelton2000temperature}. 
In this process, NIPAM monomers are dissolved in water along with crosslinking agents, most commonly N,N'-methylenebis(acrylamide) (BIS). 
Due to the differing reactivities of NIPAM monomers and BIS crosslinkers, the resulting microgels often adopt a core-shell morphology. This structure is characterized by a dense, crosslinker-rich core, surrounded by a looser, crosslinker-poor corona. 
The reference model to describe this characteristic structure is the well-established fuzzy sphere model~\cite{stiegerSmallangleNeutronScattering2004}.
NIPAM-BIS microgels are also extensively used as building blocks for more complex systems by using additional copolymers~\cite{hertleThermoresponsiveCopolymerMicrogels2013} or decorating them with functional molecules or nanoparticles~\cite{plamperFunctionalMicrogelsMicrogel2017b,karg2009smart}. The core-shell structure of these microgels is however still strongly different from that of a star polymer, resulting in markedly different properties.\cite{gury2025} 
A few studies have explored variations in the synthesis conditions in order to obtain different internal structure of microgels, in particular toward the development of more homogeneous networks~\cite{mueller2018dynamically,kyrey2019inner,stillSynthesisMicrometersizePolyNisopropylacrylamide2013}, while several recent works have addressed the case of ultra-low-crosslinked microgels~\cite{gao2003cross,bachman2015ultrasoft}, in the absence of any external crosslinking agent, but relying on self-crosslinking mechanism of NIPAM. Even in the latter case, however, the density profile is significantly different from that of star polymers.
To obtain a star-like architecture, it may be useful to use different types of crosslinkers. A pioneering study in this direction was put forward by the Hellweg group back in 2002~\cite{kratzVolumeTransitionStructure2002}, who reported the use of alternative crosslinking agents. In particular, these authors employed ethylene glycol dimethacrylate (EGDMA) instead of BIS, finding very high shrinking abilities of the particles even at large crosslinker content. They hypothesized that the fast reactivity of EGDMA, together with its tendency to self-bind, could be responsible for this peculiar behavior, however without providing detailed microscopic insights on the internal structure of the particles.
A few additional studies on the topic are available in the literature on microgels not based on PNIPAM~\cite{gawlitza2014structure,aguirre2019versatile}. More recently, in the context of copolymer microgels based on PNIPAM and poly(ethylene glycol) (PEG) with EGDMA as crosslinker, it was found that the low temperature form factors %in the presence of PEG 
could be be better described by a star polymer profile rather than by the standard fuzzy sphere model\cite{rivas2022link}.

Inspired by the possibility to establish star-like behavior in microgels, in this work we provide an extensive experimental and computational study of PNIPAM-EGDMA microgels as a function of EGDMA concentration, across the VPT. The microgels are synthesized in the presence of a fixed amount of surfactant, thereby not exceeding a few hundreds of nanometers in size. We thus complement Dynamic Light Scattering (DLS) measurements with Small Angle X-ray Scattering (SAXS) and rationalize the experimental findings with monomer-resolved numerical simulations based on two different approaches: on one hand we employ \textit{ad hoc}-designed star-like particles and on the other hand we build realistic microgels, based on the generalization of our previously established state-of-the-art method for standard PNIPAM-BIS microgels. We then put forward a new model, that relies on an appropriate combination of a star polymer colloid and a core-fuzzy-shell particle, here after named the ``star-like fuzzy sphere model'', that is able to describe the measured form factors for all crosslinker concentrations and temperatures. The analysis reveals a strong evidence of star-like morphology for PNIPAM-EGDMA microgels, particularly at low crosslinker concentration. Upon increasing the amount of EGDMA, the star-like character is retained despite the development of a distinct, crosslinker-rich core.  Our work thus provides evidence of a new class of soft nanocolloids with star-like architecture, that are obtained with a standard precipitation polymerization synthesis in water. This peculiar internal structure is still combined with the thermoresponsive character of microgels, resulting in a very sharp VPT, which is preserved at all investigated EGDMA contents. These results hold high promise for applications as well as a new model system for fundamental research, easily available by standard synthesis methods.

\section{Results and Discussion}
\subsection{Swelling Behavior of PNIPAM-EGDMA microgels}
\begin{figure}[H]
    \centering
    \includegraphics[width=1\linewidth]{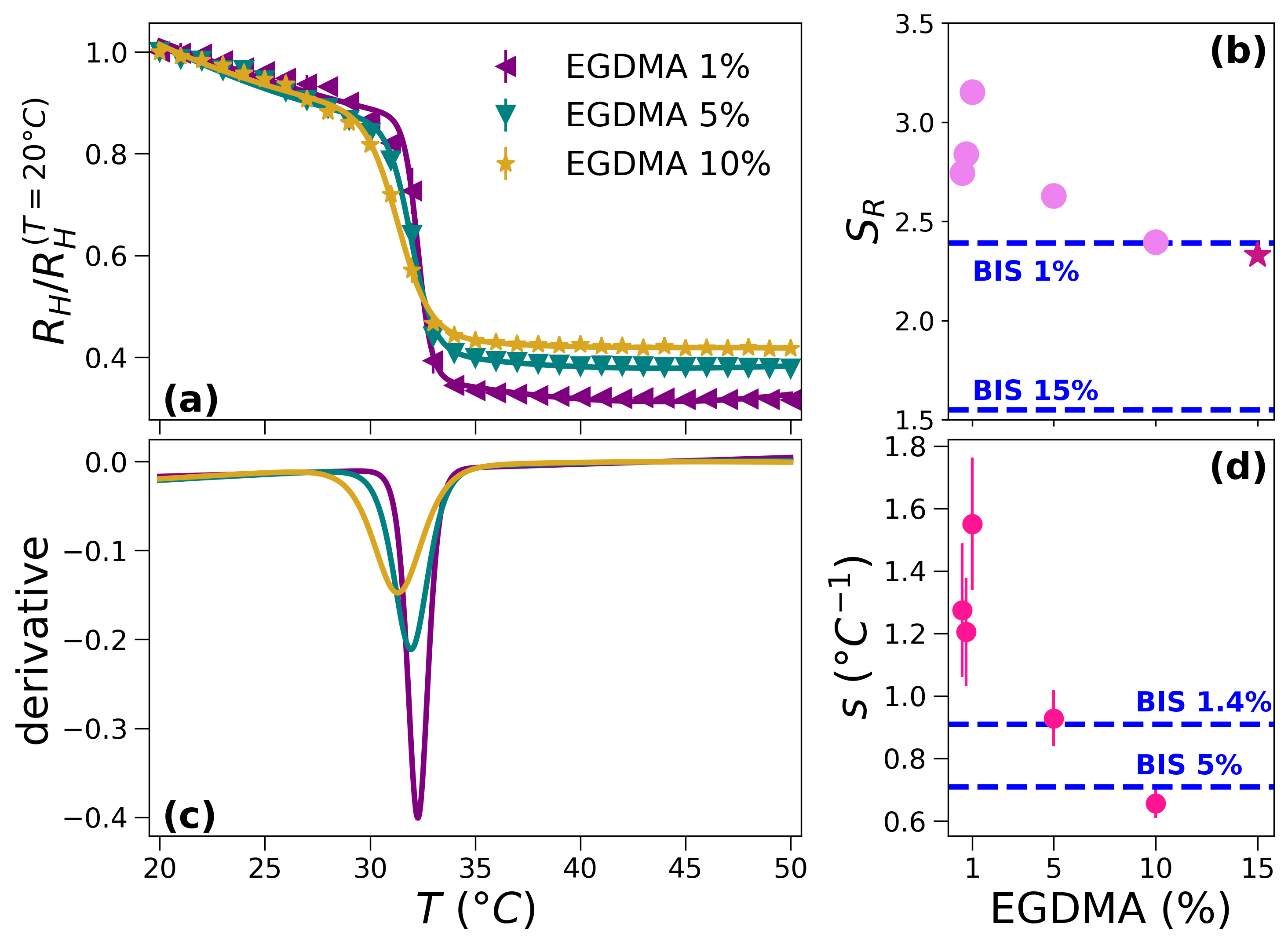}
    \caption{(a) Normalized hydrodynamic radius $R_H (T)/R_H(T=20\celsius)$ of PNIPAM-EGDMA microgels with different EGDMA content as a function of temperature $T$. For readability of the figure we report only the curves for $C_{\text{EGDMA}}=1\%$, $C_{\text{EGDMA}}=5\%$ and $C_{\text{EGDMA}}=10\%$. Solid lines are fits according to Eq.~\ref{eq:RH(T)}; (b) swelling ratio $S_R$ of the microgels as a function of crosslinker concentration. The horizontal lines represent the corresponding values for microgels synthesised at $C_{\text{BIS}}=1\%$ and $C_{\text{BIS}}=15\%$\cite{kratzVolumeTransitionStructure2002}. Violet star represents $S_R$ taken from the work of Kratz \latin{et al.}\cite{kratzVolumeTransitionStructure2002}; (c) derivative of the fits of the swelling curves reported in (a); (d) sharpness parameter $s$ extracted from the fits. The horizontal lines represent the corresponding value for microgels synthesised at $C_{\text{BIS}}=1.4\%$\cite{delmonteTwostepDeswellingVolume2021a} and $C_{\text{BIS}}=5\%$\cite{elancheliyanRoleChargeContent2022a}.}
    \label{fig:figura1}
\end{figure}

In Figure ~\ref{fig:figura1} we report the DLS characterization of the VPT of microgels synthesized with different contents of EGDMA, with molar percentages $C_{\text{EGDMA}}$ between 0.5\% and 10\%. In particular, in Figure ~\ref{fig:figura1}(a) we report the normalized hydrodynamic radius $R_H (T)/R_H(T=20\celsius)$ as a function of temperature fitted through Eq. \ref{eq:RH(T)}. For readability we show the curves only for the samples EGDMA $1\%$, $5\%$ and $10\%$ in Figure 1(a), while all additional data are reported in Fig.~\ref{fig:SI_RH} of Supporting Information (SI).
The two key features of PNIPAM-EGDMA microgels are their high swelling ability and their capacity to maintain a sharp VPT even at high crosslinker concentrations. 
The swelling ability $S_R=R_H(T=20\celsius)/R_H(T=45\celsius)$, namely the ratio between the hydrodynamic radius in the swollen and the collapsed states, is shown in Fig.~\ref{fig:figura1}(b) for all synthesized samples. 
For comparison, the dashed horizontal lines indicate the values of $S_R$ for standard PNIPAM-BIS microgels.
Notably, the swelling ratio of PNIPAM-EGDMA microgels with $C_{\text{EGDMA}}=10\%$, and even $C_{\text{EGDMA}}=15\%$~\cite{kratzVolumeTransitionStructure2002}, is comparable to the one of standard microgels with markedly lower molar percentage of crosslinker  $C_{\text{BIS}}=1\%$.
In addition, the VPT remains very sharp even at high crosslinker concentrations, as shown  directly from the swelling curves in Fig.~\ref{fig:figura1}(a) and their fitted derivatives in Fig.~\ref{fig:figura1}(c). By fitting the data to Eq.~\ref{eq:RH(T)}, we quantify the transition sharpness via the parameter $s$, that is shown as a function of $C_{\text{EGDMA}}$, in Fig.~\ref{fig:figura1}(d). By comparing the obtained values of $s$ with those reported in literature for PNIPAM-BIS microgels ($s\sim 0.9$ for c=1.4\%~\cite{delmonteTwostepDeswellingVolume2021a} and $s\sim 0.7$ for c=5\%~\cite{elancheliyanRoleChargeContent2022a}), it is clear  that the VPT for PNIPAM-EGDMA microgels occurs in a much narrower temperature range. In particular, the value of $s$ for $C_{\text{EGDMA}}=5\%$ is close to that of PNIPAM-BIS microgels with $C_{\text{BIS}}=1.4$\%~\cite{delmonteTwostepDeswellingVolume2021a}, while the one for $C_{\text{EGDMA}}=10$\% is comparable with that for $C_{\text{BIS}}=5\%$~\cite{elancheliyanRoleChargeContent2022a}. The main characteristics of PNIPAM-EGDMA microgels as determined by DLS measurements are summarized in 
Table~\ref{tab:SI_DLS_values}  of SI.

\subsection{Evidence of star-like microscopic structure of PNIPAM-EGDMA microgels}
\begin{figure}[H]
    \centering
    \includegraphics[width=1\linewidth]{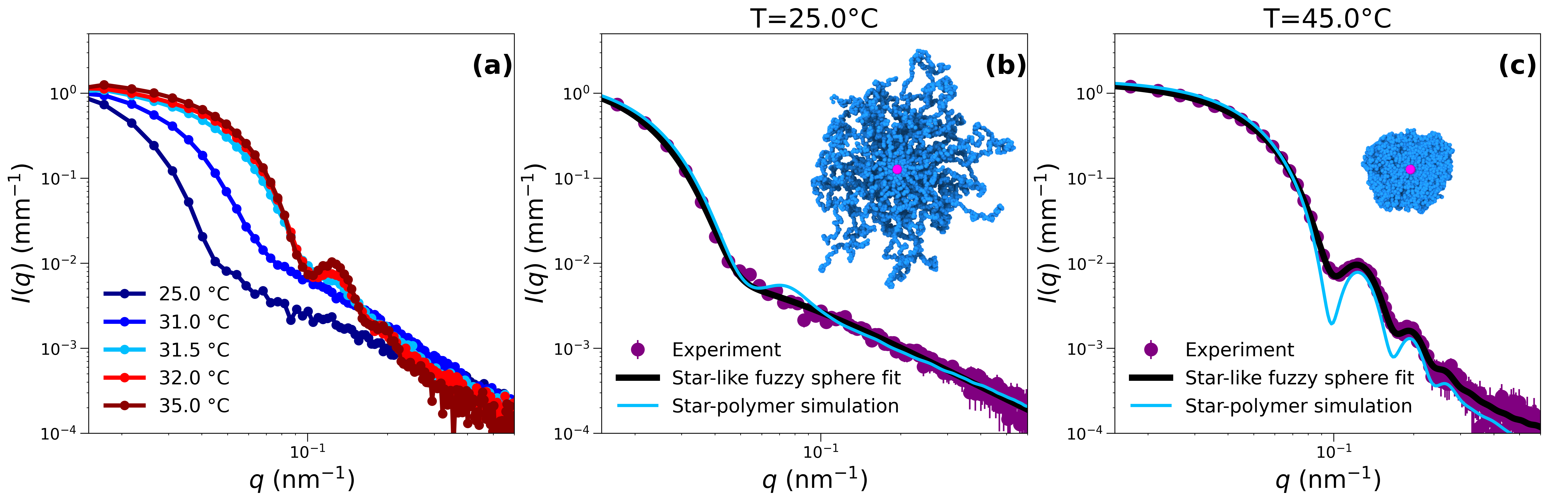}
    \caption{(a) SAXS scattering intensities $I(Q)$ for sample EGDMA 1$\%$ measured at different temperatures. Note that for $T\geq 35 \celsius$ the data no longer change; (b) and (c) measurements at $ T = 25 \celsius$ and $ T = 45 \celsius$, respectively.
    Black lines are fits via the star-like fuzzy sphere model in Eq.~\ref{eq:model}, while cyan lines are the form factors calculated for a simulated star-polymer with $R_{core} = 2.24 \,\sigma$, $f=80$, $N_f = 200$ whose snapshot and the corresponding slice at $\alpha=0.0$ (b) and $\alpha=0.8$ (c) are also shown. Here beads coloured in blue represent monomers while the core is the magenta central sphere.
    }
    \label{fig:figura2}
\end{figure}

To shed light on the microscopic structure, giving rise to the particular trend of the hydrodynamic radius as a function of temperature, we performed SAXS measurements. Figure~\ref{fig:figura2}(a) shows the form factors for microgels with $C_{\text{EGDMA}}=1\%$ as a function of temperature. We can see that at low $T$ the form factors display a peak-less shape, that is maintained up to the VPT, where suddenly there is a change of structure, occurring essentially between $31\celsius$   and  $32\celsius$. For higher temperatures $T\geq 35\celsius$ no further change of the form factors is detected in the whole investigated range. These results confirm the sharpness of the transition observed by DLS.

Next we focus on the form factors at two representative temperatures, below and above the VPT, shown in Fig.~\ref{fig:figura2}(b) and (c). Building on the hypothesis of Kratz and coworkers~\cite{kratzVolumeTransitionStructure2002} that there is an accumulation of crosslinkers in the core, we test a star-like morphology for our microgels. This is done in two ways: (i) we introduce a novel star-like fuzzy sphere model to describe the experimental form factors and (ii) we simulate star polymers with varying number of arms and core properties to compare with experimental data.

We start by discussing the model, defined in 
Eq.~\ref{eq:model} and described in detail in Model section, that we developed in order to describe the peculiar structure of star-like microgels. In the model, the form factor is expressed as the sum of a term $P_{mgel}$ (Eq.~\ref{eq:core-fuzzy-shell_model}), which describes the overall structure of the particle, and second term $P_{star}$ (Eq.~\ref{eq:star_model}), already put forward by Dozier and coworkers~\cite{dozierColloidalNatureStar1991}, accounting for the short-range correlations inside the star corona. In particular, $P_{mgel}$ is the form factor of a particle with a core complemented by a fuzzy shell that is needed in order to incorporate the mixed nature of star-like microgels, including both the fuzzy structure of microgels and a core region that can become more extended by increasing the amount of crosslinker. 
We find that Eq.~\ref{eq:model} is able to quantitatively describe the experimental form factors of PNIPAM-EGDMA microgels with $C_{\text{EGDMA}}=1\%$, as shown in Fig.~\ref{fig:figura2}(b) and (c). Fit parameters for the two reported temperatures are summarized in Table \ref{tab:par_coreshellfuzzy}.
In particular, at $T=25\celsius$ the fit yields a rather small core radius, $R_c=6$ nm, shell thickness $t=48$ nm and fuzziness $\sigma_s=47$ nm. The exponent $\mu$ is found to be $\approx 0.66$,  in light with expectations for good solvent conditions $(\nu \simeq 0.588)$. The total radius of the microgels, $R_t=R_c+t+2\,\sigma_s$, is found to be consistent with the hydrodynamic radius $R_H(T=25\celsius)=156$nm estimated by DLS and reported in Table \ref{tab:SI_DLS_values}. Indeed, we find $R_t(T=25\celsius)=148$ nm. From the fit, we observe slight differences in scattering length density between the shell and solvent ($\Delta\rho^{s0}$) and between the core and shell ($\Delta\rho^{cs}$), consistent with the microgel being fully solvated.
From the low temperature form factors, it is not possible to estimate the polydispersity of the samples, since the model reported in Eq.~\ref{eq:model}  is always able to fit the data independently of polydispersity. We also note that at low temperature an equally good fit is found also with the standard model for colloidal star polymers~\cite{dozierColloidalNatureStar1991}, as shown in Fig.~\ref{fig:SI_EGDMA1_APS}(a). This indicates an internal structure that is fully compatible with that of a true star polymer.  
However, since to describe the behavior at high temperatures the star model is not sufficient, we rely on the star-like fuzzy sphere model that works for all investigated conditions.
Indeed, at high temperature the microgel collapses and its shape becomes spherical.
By fitting the form factor at $T=45\celsius$ through Eq.~\ref{eq:model}, we find $R_c=38$ nm, $t=9$ nm and $\sigma_s=3$ nm giving a total radius $R_t=53$ nm, again in agreement with $R_H(T=45\celsius)=51.3$nm (Tab.~\ref{tab:SI_DLS_values}), while the value $\mu \approx 1.7$ signals the worsening of the solvent conditions. 
Given the peculiarity of the obtained structure, we test its robustness by using a different initiator in the synthesis.
To this aim, we report in the SI the results for a PNIPAM-EGDMA microgel synthesized using the same fraction of EGDMA, by using APS rather than KPS as initiator.  
Figure \ref{fig:SI_EGDMA1_APS}(b) clearly shows that, also in the presence of APS, a star-like behavior is observed and the data are again well described by the star-like fuzzy sphere model.

In order to validate the star architecture of the microgels at low $T$, we further compare the results with those calculated in the simulations of a star polymer, following the protocol described in Methods. The \textit{in silico} star that best reproduces the experimental microgel with $C_{\text{EGDMA}}=1\%$, also reported in Fig.~\ref{fig:figura2}, is found to have a core radius $R_c = 2.24\,\sigma$, arm number $f = 80$ and number of monomers per arm $N_f = 200$. These parameters fullfill the condition where the number of arms is able to fully cover the core, as in the standard blob description of a star~\cite{likos2006soft}. The model is capable to reproduce the deswelling behavior of the microgel by increasing temperature, using a value of the solvophobic parameter $\alpha=0.8$ in the collapsed state (see Methods), that is the same $\alpha$ value that is normally employed to fit corresponding data for PNIPAM-BIS microgels~\cite{ninarello2019modeling,hazra2023structure}.  The snapshots of the star that best represents the microgel are also shown in Fig.~\ref{fig:figura2}, denoting the presence of a tiny finite core, that is responsible for the small peak present in the numerical form factor at $\alpha=0.0$.

\begin{figure}[H]
    \centering
    \includegraphics[width=1\linewidth]{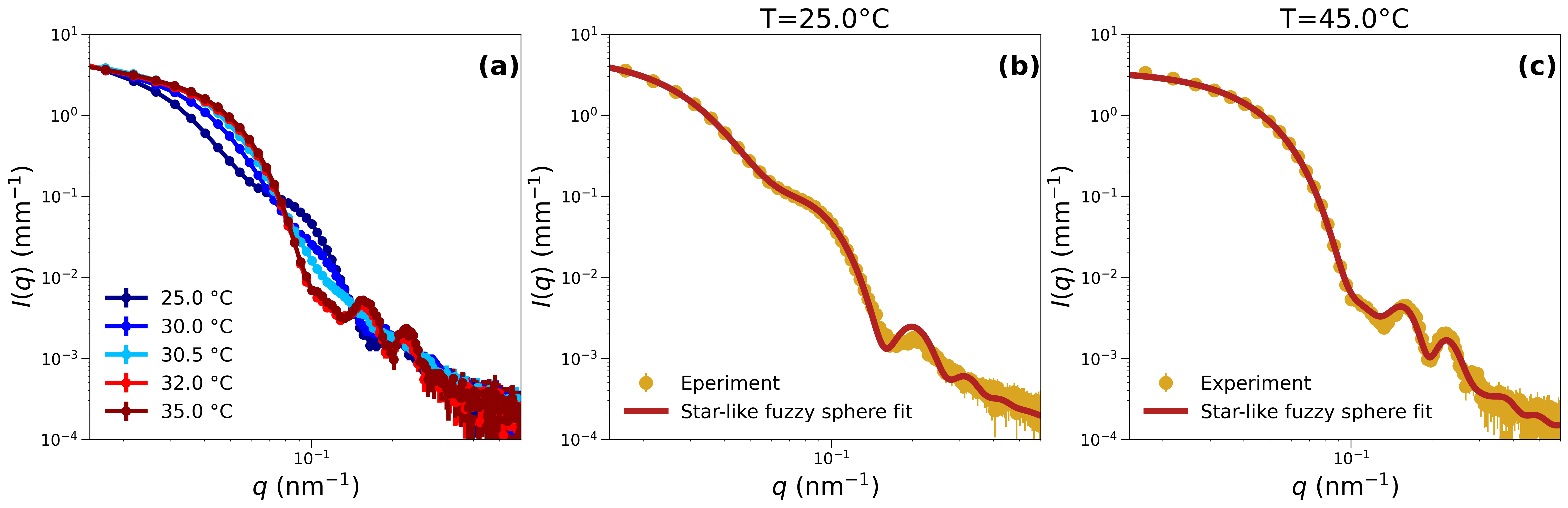}
    \caption{SAXS scattering intensities $I(q)$ for sample with $C_{\text{EGDMA}}=10\%$ measured at different temperatures (a). Again, note that for $T\geq 35 \celsius$ the data no longer change. In (b) and (c) focus on the measurements at $ T = 25 \celsius$ and $ T = 45 \celsius$. Brown solid lines represent fits in which the form factor P(q) is represented by the star-like fuzzy sphere model presented in Eq.~\ref{eq:model}.  }
    \label{fig:figura3}
\end{figure}

Next, we move on to examine the experimental form factors of microgels with $C_{\text{EGDMA}}=10\%$. These are reported in Fig.~\ref{fig:figura3}(a) at different temperatures. Here a more complex structural change is observed at low $T$, as compared to Fig.~\ref{fig:figura2}, but still the microgel is found to sharply collapse around $T=32\celsius$, and for $T\geq 35\celsius$ its structure does not change any longer. We then focus on the SAXS data acquired at low and high temperatures, shown in Fig.~\ref{fig:figura3}(b) and (c) respectively, together with the fits performed using the star-like fuzzy sphere  model of Eq.~\ref{eq:model}, which is found also in this case to quantitatively describe the measured data at both temperatures.
The resulting fit parameters are reported in  Table~\ref{tab:par_coreshellfuzzy}, showing that the size of the inner core $R_c$ remains roughly constant upon varying the temperature. In particular, we find $R_c(T=25\celsius)=28$ nm and $R_c(T=45\celsius)=25$ nm. This is the signature of the presence of a stiff central region with a large density of crosslinkers, that does not strongly respond to temperature changes. Instead, a significant decrease is observed for both the shell thickness and the fuzziness, by increasing $T$. The total size of the microgels is found to be consistent at both temperatures with the hydrodynamic radius $R_H(T=25\celsius)=141.7$ nm and $R_H(T=45\celsius)=62.5$ nm as reported in Table ~\ref{tab:SI_DLS_values}. 
Indeed, we find $R_t(T=25\celsius)=129$ nm and $R_t(T=45\celsius)=67$ nm.
From the fit we obtain a polydispersity of $7\%$ in $R_c$ at low temperature; while at high $T$ the model fits the data very well without including polydispersity. 
As done in the case $C_{\text{EGDMA}}=1\%$, we tried to compare the experimental form factors with those calculated in the simulations of a star polymer (Fig.~\ref{fig:SI_EGDMA10_starpolymer}). In this case, since EGDMA forms a larger central core, the simulated star polymer fails to describe the experimental results, as detailed in the SI.
Hence, we can only capture the qualitative behavior at low-$T$ but not the  one at high-$T$. Altogether these results suggest that, as $C_{\text{EGDMA}}$ increases, the resulting structure at low temperatures differs from that of a simple star polymer, becoming more similar to a core-fuzzy shell. 
This clearly indicates a mixture between a star-like object with a central core and a microgel particle with a fuzzy density profile, establishing the star-like fuzzy sphere model of Eq.~\ref{eq:model} as the most appropriate to correctly capture the internal structure of PNIPAM-EGDMA microgels, independently of the employed EGDMA amount.

\begin{table}[H]
    \centering
    \resizebox{\textwidth}{!}{
    \begin{tabular}{|ccccccccc|}
    \hline
      EGDMA (\%) & $T$ ($\celsius$) & $R_c$ (nm) & $t$ (nm)  & $\sigma_s$ (nm)& $\mu$ & $\xi$ (nm)&$\Delta\rho^{cs}$ (nm$^{-2}$)&$\Delta\rho^{s0}$ (nm$^{-2}$)\\ \hline
      1 & 25 & 6 & 48 & 47 & 0.66 & 21 &0.03 &0.01 \\
      1 & 45 & 38 & 9 & 3 & 1.74 &28  &0.02 &0.03 \\
      10 & 25 & 28 &51 & 25 &0.66 &20 &0.03 & 0.001\\
      10 & 45 & 25 &28 & 7&1.60 & 28& 0.008 &0.003 \\
      \hline
    \end{tabular}
    }
            \caption{Best-fit parameters derived from the star-like fuzzy sphere model (Eq.~\ref{eq:model}) for samples EGDMA $1\%$ and EGDMA $10\%$. $\Delta\rho^{cs}$ and $\Delta\rho^{s0}$ represent the differences of the scattering length densities between the core and the shell ($\rho_c-\rho_s$) and between the shell and the solvent ($\rho_s-\rho_0$), respectively. Errors are omitted for visual clarity of the table.}
    \label{tab:par_coreshellfuzzy}
\end{table}

\subsection{Monomer-resolved simulations of star-like microgels}
To get microscopic insights on this behavior, we build on our previous works on PNIPAM-BIS microgels and develop a new monomer-resolved microgel model, able to capture the PNIPAM-EGDMA architecture. As described in Methods, we generalize our recipe put forward in Refs.~\citenum{ninarello2019modeling} and \citenum{hazra2023structure} to take into account the extremely fast polymerization kinetics of EGDMA and the tendency of the crosslinking molecules to bind among themselves. In this way, we are able to reproduce the full form factor experimental behavior, upon varying $C_{\text{EGDMA}}$  and temperature, with a consistent set of parameters that uses the same nominal $C_{\text{EGDMA}}$ as in experiments. 

 \begin{figure}[H]
    \centering
    \includegraphics[width=1\linewidth]{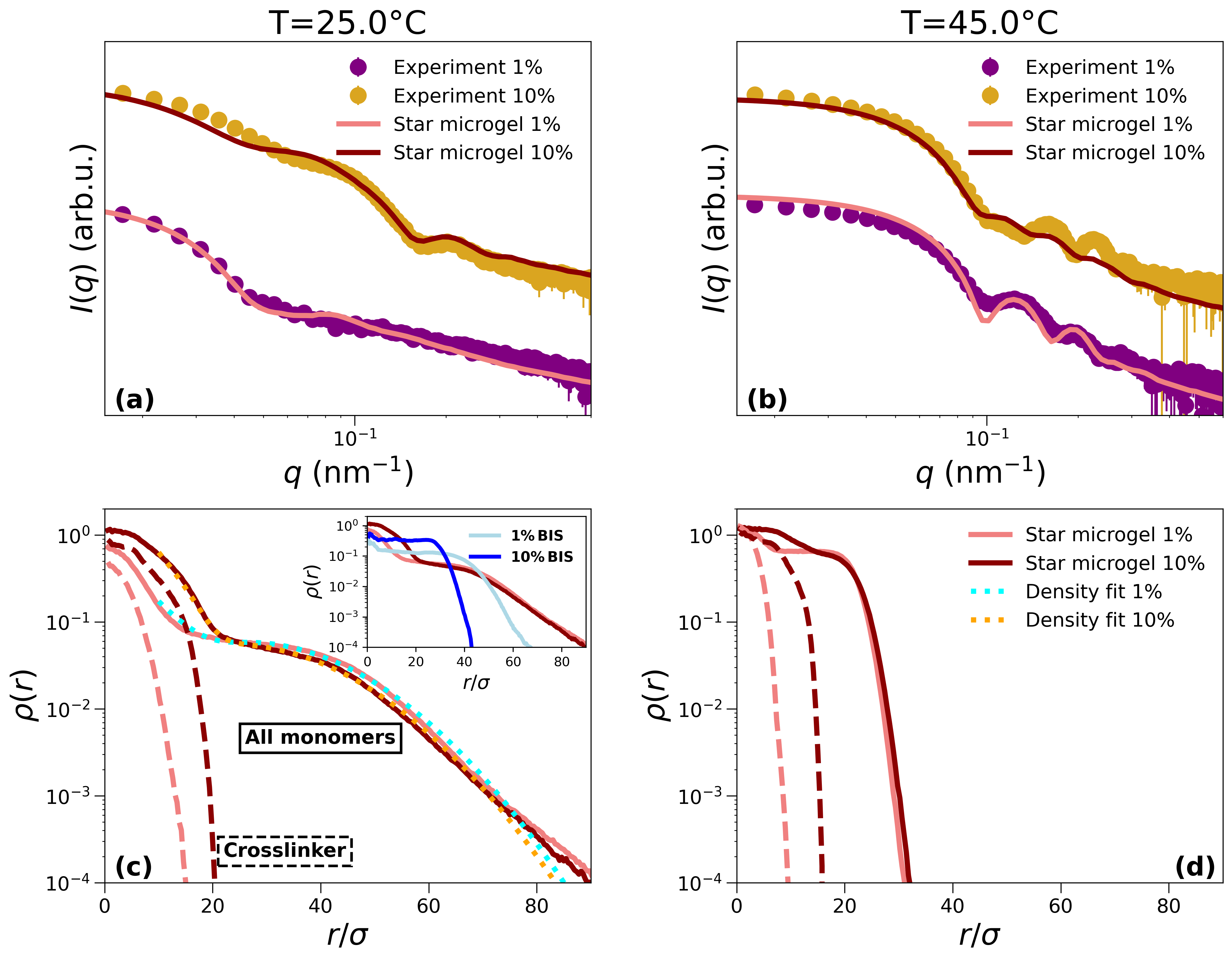}
    \caption{SAXS scattering intensities $I(Q)$ for PNIPAM-EGDMA microgels with   $C_{\text{EGDMA}}=1\%$ and $C_{\text{EGDMA}}=10\%$ at $T = 25\celsius$ (a) and $T = 45\celsius$ (b). The curves are vertically shifted for clarity.
    Solid lines represent the form factors of  simulated star-microgels at $\alpha=0.0$ (a) and $\alpha=0.8$ (b). Numerical curves are superimposed onto experimental ones by rescaling for the monomer size $\sigma=0.48$ nm and 0.43 nm, respectively for $C_{\text{EGDMA}}=1\%$ and $C_{\text{EGDMA}}=10\%$. In (c) and (d) we show the density profiles of the star-microgels at $\alpha = 0.0$  and $\alpha = 0.8$ respectively as a function of the distance from the center of mass, rescaled by the monomer size. Density profiles at $\alpha=0.0$ are well-described by Eq.~\ref{eq:Rho_Star}, representative of star behavior, shown as dotted lines. Dashed lines are the density profiles of the crosslinkers only. The inset in (c)  shows the comparison between density profiles of PNIPAM-BIS and PNIPAM-EGDMA microgels.
    }
    \label{fig:figura4}
\end{figure}
\begin{figure}[H]
    \centering
    \includegraphics[width=1\linewidth]{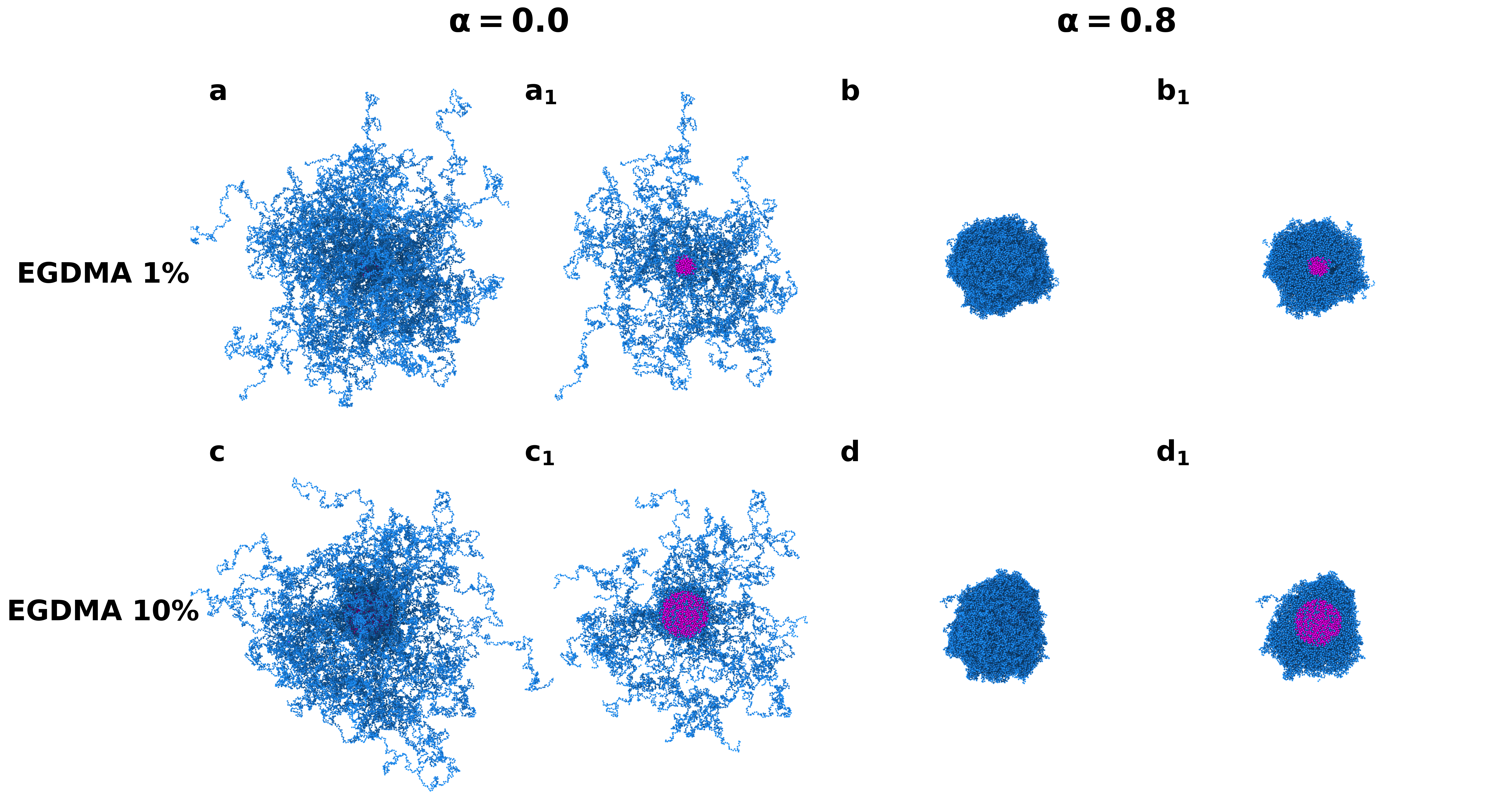}
    \caption{Snapshots of simulated PNIPAM-EGDMA star-like microgels with  $C_{\text{EGDMA}}=1\%$ and $C_{\text{EGDMA}}=10\%$ at $\alpha=0.0$ (a,c) and $\alpha=0.8$ (b,d). The corresponding slice is shown next to each snapshot to highlight the crosslinkers, coloured in magenta, for EGDMA $1\%$ (a$_1$)(b$_1$) and for EGDMA $10\%$ (c$_1$)(d$_1$). Note that the core is made by crosslinkers as well as a significant fraction of monomers for the $C_{\text{EGDMA}}=10$\% case. 
    }
    \label{fig:figura5}
\end{figure}

The comparison between the calculated form factors for the new {\it in silico} PNIPAM-EGDMA microgels and experiments is reported in Fig.~\ref{fig:figura4}, again for the two representative temperatures $ T = 25 \celsius$ and $ T = 45 \celsius$, respectively in panels (a) and (b), for both $C_{\text{EGDMA}}=1\%$ and $C_{\text{EGDMA}}=10\%$. The agreement between simulations and experiments is quite accurate. This is obtained by simply rescaling the simulation data onto the experimental ones, as done in previous works~\cite{ninarello2019modeling}, using the monomer size $\sigma=0.48$ nm and $0.43$ nm, respectively for the two microgels. These values are then maintained at all temperatures. We find that the high-$T$ data are well-captured using an effective temperature $\alpha=0.8$, consistently for the star polymer model in Fig.~\ref{fig:figura2} and for PNIPAM-BIS microgels~\cite{ninarello2019modeling,hazra2023structure}. These results highlight the robustness of our {\it in silico} microgel preparation, making it able to reproduce microgels with very diverse architectures. Furthermore, the availability of a microscopic model enables us to look at the internal structure of the PNIPAM-EGDMA microgels in real space. 
We thus report the radial density profiles of the microgels in Fig. \ref{fig:figura4}(c) and (d), respectively for low and high temperatures. Together with the full profiles, we also isolate the density profile of the crosslinkers, also reported in the Figures.
They do indicate that the crosslinkers are clearly localized in a tiny region of space, the particles core, whose size changes very little with increasing temperature. These findings indicate that the chains departing from the core can thus be considered as true isolated arms, as in star polymers. This is confirmed by the snapshots of the microgels, and the corresponding slices to highlight the crosslinker-rich core, reported in Fig.~\ref{fig:figura5}. It is clear that for $C_{\text{EGDMA}}=1\%$ the core is entirely made of crosslinkers and no crosslinkers are outside of it. However, it is also evident that the arms are widely polydisperse in length and we also find that sometimes they even connect from one point to another on the core. Estimating the core radius as $R_c\sim 5.7\,\sigma$ from a Gaussian fit of the crosslinker density profile, we obtain that it is quite small in comparison to the total microgel size and that its inner packing fraction is $\sim 0.28$. We then count the number of arms departing from the core, getting  approximately $f\sim 200$  for $C_{\text{EGDMA}}=1\%$ and a surface coverage fraction~\cite{likos1998star} $\gamma=f/16r^2_c \sim 0.38$.
Although the core coverage  is only partial, differently  from the case of an ideal star polymer model, our microgel still shows a density profile that is quite compatible with the star polymer predictions at large distances (see Methods, Eq.~\ref{eq:Rho_Star}), also reported in Fig.~\ref{fig:figura4}(c). This is found to be valid also for the more crosslinked microgel. Clearly, the density profiles of the microgels with both concentrations of EGDMA are very different from the corresponding ones obtained for PNIPAM-BIS microgels, also reported in the inset of Fig. \ref{fig:figura4}(c), indicating a very different structure also at high EGDMA content. In this case, the core becomes much more extended, exceeding the scaling behavior with $\sim C_{\text{EGDMA}}\ ^{1/3}$, since we observe from the snapshots in Fig.~\ref{fig:figura5} that a significant fraction of monomers also remains trapped in the core. Therefore, in this case, it is not obvious to identify a clear star structure and also to count the arms departing from the core. We thus believe that the additional monomers belonging to the core give rise to the fuzziness of particles, that is needed to describe the experimental data in the star-like fuzzy sphere model. 
At high temperature, the simulations fully capture the evolution of the peak positions, including the depletion of the first peak observed for $C_{\text{EGDMA}}=10\%$, that is quite reminiscent of that observed in core-shell microgels where the core is a solid silica core~\cite{hildebrandt2023fluid}, again reinforcing the choice of our star-like fuzzy sphere model. We note that the peaks are less pronounced in the simulations, probably due to finite size effects, as increasing the number of monomers is known to increase the high-$q$ oscillations~\cite{ninarello2019modeling}.
We also report in SI (Figure \ref{fig:SI_densityprofile}) the comparison between the density profiles of EGDMA $1\%$ star-microgel and EGDMA $1\%$ star-polymer, to show that they are quite similar to each other and  also compatible with Eq.~\ref{eq:Rho_Star}.  Altogether these simulations show that, even if our microgels are built from the self-assembly of monomers and crosslinkers mixed in the same stoichiometric ratio as in experiments and not as designed stars, we do recover a star-like structure that is mixed with microgel features for these new fascinating soft particles.

\begin{figure}[H]
    \centering
    \includegraphics[width=1\linewidth]{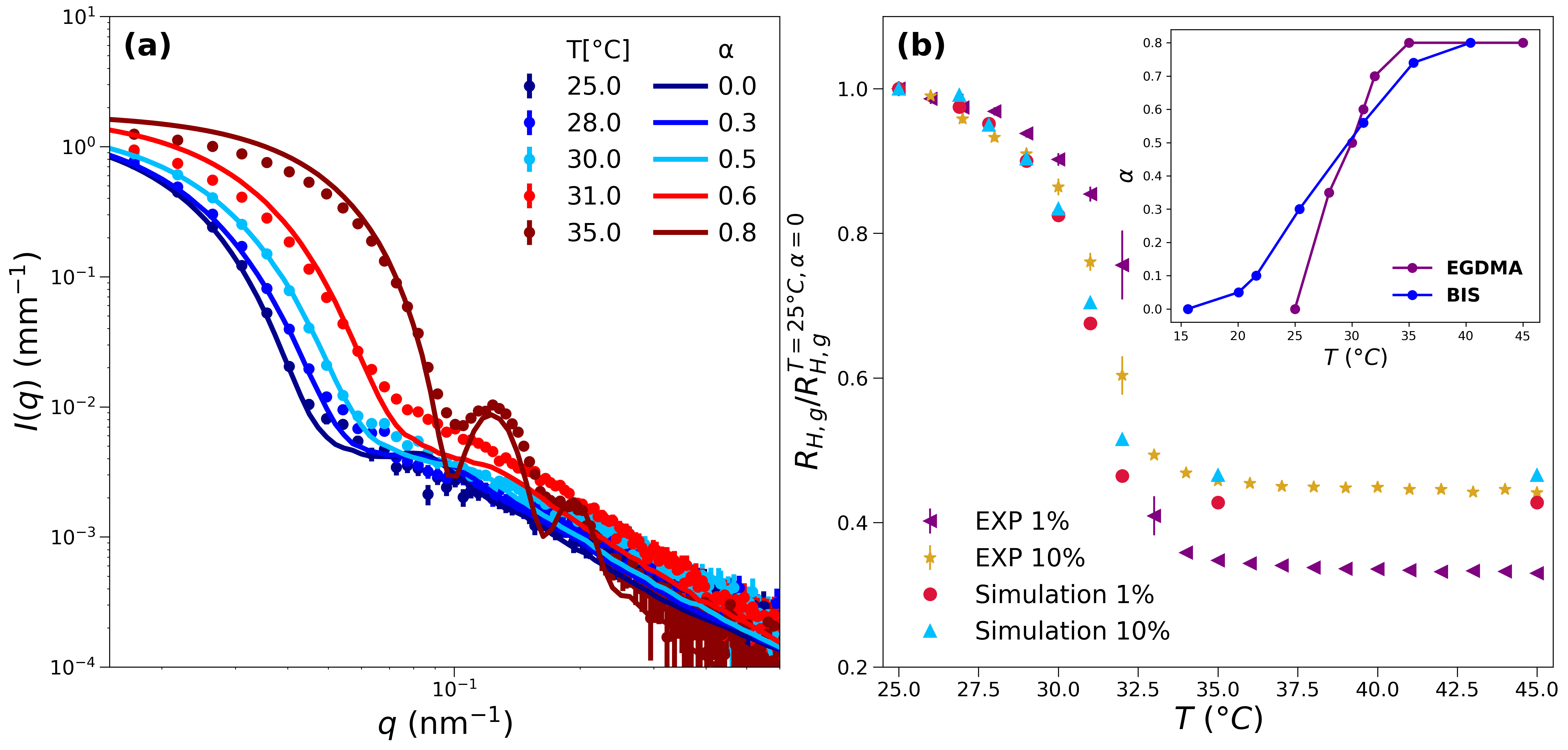}
    \caption{ (a) Comparison between experimental (full symbols) and numerical form factors (solid lines) of PNIPAM-EGDMA microgels with $C_{EGDMA}=1\%$. The numerical form factors are rescaled along the x-axis  by the same factor $\sigma=0.48$nm at all temperatures. (b) Comparison between the hydrodynamic radius measured with DLS and radius of gyration calculated from simulation as a function of temperature. The mapping  between the solvophobic parameter $\alpha$ and $T$ is shown in the inset and compared with the one of PNIPAM-BIS microgels~\cite{ninarello2019modeling}. 
    }
    \label{fig:figura6}
\end{figure}

To further analyse the star-like microgels synthesized with EGDMA, we now proceed with a detailed comparison of the experimental form factors with those calculated with the simulated microgels in the full temperature range. In this way we can establish a mapping between the temperature in $\celsius$ and the solvophobic parameter $\alpha$, similarly to what found for PNIPAM-BIS microgels~\cite{ninarello2019modeling}. In Figure \ref{fig:figura6} (a) we compare numerical and experimental $P(q)$ for $C_{\text{EGDMA}}=1\%$ for several values of temperature and the corresponding solvophobic parameter $\alpha$.
The resulting mapping between $\alpha$ and T, also found to be valid for $C_{\text{EGDMA}}=10\%$ microgels, is reported in the inset of Figure \ref{fig:figura6} (b) together with the same mapping done for standard PNIPAM-BIS microgels by Ninarello et al.\cite{ninarello2019modeling}. Curiously, we find that while the VPT temperature remains the same, the two types of microgels explore the temperature range with a rather different effective solvophobic potential. This is due to the absence of crosslinkers in the outer region of the stars, which makes the transition from swollen to collapse rather sharp, in comparison to standard microgels, where the presence of crosslinkers in the corona gives rise to a more cooperative transition.
Additionally, in Fig.~\ref{fig:SI_Rg_alpha} we show the comparison between the normalized radius of gyration between \textit{in silico} PNIPAM-BIS and PNIPAM-EGDMA microgels to note once again that with the model introduced in the simulations we are able to obtain star-like microgels that have a high swelling capacity and a very sharp VPT.

\section{Conclusions}
In this work we presented an extensive characterization by experiments and simulations of a new class of soft nanocolloids: PNIPAM-EGDMA microgels. Despite they were first reported in 2002~\cite{kratzVolumeTransitionStructure2002} and sometimes EGDMA is used to replace BIS as crosslinker in studies of co-polymerized microgels~\cite{serrano2012synthesis,clara-raholaStructuralPropertiesThermoresponsive2012,sayed2019pegylated,rivas2022link,ruiz-francoConcentrationTemperatureDependent2023c}, there was not yet a dedicated study devoted to unveil their peculiar inner structure. Building on the fascinating hypothesis that EGDMA could concentrate in the middle of the microgels due to its fast reactivity, we have measured their form factors with high-brilliance Small Angle X-Ray Scattering at ESRF and compared them with state-of-the-art simulations to finally reveal a quite complex architecture.

Indeed, we demonstrate that PNIPAM-EGDMA microgels are structurally very similar to ideal star polymers for low crosslinker concentration. This opens the possibility to investigate these very intriguing systems, largely studied in the past theoretically and also experimentally, whose spreading in the last years has been hindered due to their difficult synthesis. The use of a simple polymerization reaction is indeed a major breakthrough to their exploitation, particularly to investigate dynamical behavior and rheology of dense suspensions~\cite{vlassopoulos2014tunable,van2017fragility}. Of course, we have now shown that the number of arms is set by the synthesis in the order of $\sim 10^2$, but we expect that modifications of the simple batch approach, e.g. semi-batch, could be employed to vary the number of arms and to tune the properties of the stars.
Furthermore, upon increasing EGDMA concentration, we have observed that the microgels acquire a larger core, but still maintain arms that are basically non-crosslinked to each other, which would be also interesting to compare to polymer-grafted nanoparticles~\cite{gury2021internal}. Here, the peculiarity of the present soft particles is that the core is not solid, but also maintains some degree of thermoresponsiveness and with some degree of fuzziness, as shown by our form factor description. This hybrid nature of the microgels makes them very appealing for applications in different fields, as they bridge properties of particles that were classified in different classes in the soft particle catalogue, but here they can be tuned simply changing EGDMA content and perhaps introducing new parameters to the synthesis, e.g. additional copolymers. It will also be important to classify the softness of these new particles in the context of existing ones~\cite{vlassopoulos2014tunable,scotti2022softness}.

To characterize the unusual structure of star-like microgels we have introduced a new model, combining star polymer and microgel features, which is able to describe the form factors of the microgels at all temperatures and EGDMA concentrations. This model also reveals the strong star-like character of the microgels at low crosslinker concentration, that are very similar to true star polymers, as also demonstrated by numerical simulations. In addition, we have shown that our monomer-resolved PNIPAM-EGDMA microgel model is able to describe the experimental samples with a consistent set of parameters, using very simple and intuitive modifications of our previously established assembly of PNIPAM-BIS microgels. Specifically, we modified the range and the strength of the force acting on the crosslinkers during assembly in order to mimic the faster reactivity of EGDMA as compared to BIS, and we then varied the range of this force consistently with the concentration of EGDMA. These findings on one hand reinforce the potential of the {\it in silico} synthesis, which can be adapted to describe microgels of virtually any structure, and on the other hand make it possible to further develop in the future the investigation of these microgels also under different conditions, e.g. at interfaces~\cite{camerin2019microgels,menath2021defined} or in dense suspensions~\cite{nikolov2020behavior,del2024numerical}, as already done for PNIPAM-BIS microgels. In addition, the theoretical characterization of these new microgels will be important and will have to be compared to the logarithmic effective potential prediction~\cite{likos1998star} that holds for star polymers. Potentially, our investigation will open a new perspective in the study of soft nanocolloids, addressing  the physical interplay  between elastic (microgels) and ultrasoft (stars) interactions.

Finally, from the point of view of applications, it is important to stress that PNIPAM-EGDMA microgels provide star-like soft particles with inherent thermoresponsivity, a novel feature giving rise to the occurrence of a VPT that is much sharper than in standard PNIPAM-BIS microgels. This behavior, already observed in the pioneering work of Hellweg and coworkers~\cite{kratzVolumeTransitionStructure2002}, is now rationalized by the star-like architecture, which allows a large freedom to the arms, thus being able to experience more suddenly the increased hydrophobic interactions with the solvent with respect to a crosslinked corona. The obtained sharp transition can be of course very appealing for all those applications requiring a sudden volume change of the microgels, e.g. sensing or drug delivery.

\section*{Experimental Methods}

\subsection*{Chemicals}
The N-isopropylacrylamide monomer (NIPAM) (Sigma-Aldrich) was recrystallized from hexane and methanol, and then dried under reduced pressure (0.01 mmHg) at room temperature. Ethylene glycol dimethacrylate (EGDMA), Sodium dodecyl sulfate (SDS) with 98\% purity, potassium persulfate (KPS) and ammonium persulfate (APS) (all purchased from Sigma-Aldrich) were used as received. 

\subsection*{Synthesis of the Microgels}

Microgels with different molar percentages of crosslinker EGDMA (from 0.5\% to 10\%) were prepared under precipitation method conditions.
Specifically, NIPAM monomers, crosslinker EGDMA and the surfactant SDS were dissolved in 26.5 mL of MilliQ water and placed in a 50 mL two-necked reactor, equipped with a condenser and magnetic stirrer. The reactor is immersed in an oil thermal bath whose temperature is controlled through a heating plate.
The solution was purged under a nitrogen stream for 1 hour at room temperature. Subsequently, the temperature was raised to 70$\celsius$ and then polymerization was initiated by a controlled addition of KPS (11.6 mg dissolved in 1.2 mL of deoxygenated water) at a rate of 1 mL/min. The reaction was carried out at 70$\celsius$ for 5 hours. The resulting PNIPAM microgels were purified by dialysis in a cellulose membrane (MWCO: 6-8 kD) against ultrapure water for two weeks, with water changes twice daily. The microgel solutions were then freeze-dried and stored in the dark at 4$\celsius$.
The exact amounts of reagents used in the different syntheses are provided in Table \ref{tab:reagents}.

\begin{table}[!ht]
    \centering
    \begin{tabular}{|lccc|}
         \hline
         Sample&   EGDMA (\textmu L)&NIPAM (mg) & SDS (mg)\\
         \hline
         EGDMA 0.5\%  &      4.16 & 497.22 & 12.7\\
         EGDMA 0.7\%  &      5.83 & 496.22 & 12.7\\
         EGDMA 1\%  &      8.30 & 494.7 & 12.7\\
         EGDMA 5\%& 41.60& 474.7& 12.7\\
         EGDMA 10\%& 83.30& 449.7 & 12.7 \\
         \hline
         \end{tabular}
         
    \caption{Quantities of reagents used for syntheses of  microgels.}
    \label{tab:reagents}
\end{table}

\subsection*{Dynamic Light Scattering}
The hydrodynamic radius, $R_H$, was measured as a function of temperature using Dynamic Light Scattering (DLS). Measurements were performed with a NanoZetaSizer apparatus (Malvern Instruments LTD.) equipped with a He-Ne laser (5 mW power, 633 nm wavelength) that collects light at an angle of 173$\degree$.  
Measurements were carried out diluting the samples to a concentration of C = 0.01 wt$\%$ in MilliQ water in the temperature range between $20 \celsius$ and $50 \celsius$. After each temperature change, we equilibrated the  sample for 5 minutes.  Hydrodynamic radius $R_H$ and polydispersity index were determined by cumulant analysis \cite{koppel1972analysis}.

To describe the swelling behavior, we fit the hydrodynamic radius versus temperature with the function~\cite{delmonteTwostepDeswellingVolume2021a}:
\begin{equation}
    R_H(T)=R_0 - \Delta R\tanh{(s(T-T_c))}+A(T-T_c)+A_1(T-T_c)^2+A_2(T-T_c)^3
    \label{eq:RH(T)}
\end{equation}
where $R_0$ is the radius of the microgel at the VPT, $T_c$ is the VPT  temperature, $\Delta R$ is the amplitude of the VPT and the parameter $s$ quantifies its sharpness. The function includes a third-order polynomial inserted to describe well the trend of the hydrodynamic radius over the entire temperature range even far from $T_c$.
To characterize the VPT, we also evaluate the swelling ratio $S_R=R_H(T=20 \celsius)/R_H(T=45 \celsius)$.

\subsection*{Small Angle X-ray Scattering}
Small Angle X-ray Scattering (SAXS) experiments were performed at the ID02 beamline of the European Synchrotron Radiation Facility (ESRF) \cite{narayanan2022}. 
Samples were measured at low concentration of 0.1 wt\%, that ensures direct measuring of the microgel form factor.
Indeed, the SAXS scattered intensities can be expressed as:
\begin{equation}
    I(q)= \phi V(\Delta\rho)^2P(q)S(q)
\end{equation}
where $\phi$ is the particle volume fraction, V is the particle volume, $\Delta\rho$ the scattering length density difference between the microgels and the solvent, $P(q)$ the particle form factor, and $S(Q)$ the structure factor. Since we performed measurements in diluted condition the contribution of the structure factor can be neglected; thus the measured curves are directly linked to the form factor through a multiplicative prefactor.
For measurements, samples were placed in quartz capillaries (2mm diameter). 
We acquired SAXS curves using a Eiger2X 4M pixel detector, at temperatures selected between $25\celsius$ and $45 \celsius$ using a Peltier stage; after each temperature change, samples were equilibrated for 5 minutes before measurements. The sample-detector distance was set to 5 m in order to achieve a $q$-range of $1 \cdot 10^{-2} \, \text{nm}^{-1} \leq q \leq 1 \, \text{nm}^{-1}$ where $q$ is defined as $q=(4\pi/\lambda)\sin\theta$, $2\theta$ is the scattering angle, and $\lambda$ is the wavelength of the radiation. The exposure time for acquisitions was set to 0.5 s and 10 scattering patterns were acquired for each temperature.
Scattering patterns of a capillary filled with water were recorded for background subtraction.
The processing and averaging of the scattering patterns were performed by the software SAXSutilities~\cite{sztucki2021SAXSutilities2}.
When averaging, any scattering curve not perfectly superimposed with the overall set acquired, due to possible residual equilibration or other experimental perturbations, was discarded. Curve fitting was carried out using the software SasView~\cite{SasView}, employing the model for particle form factors introduced in the following section.

\subsection*{Structural model and form factor of star microgels}
We introduce a model that is able to describe the internal structure of star-like microgel particles and we derive an analytical expression of their form factor. The particles are composed by a highly crosslinked core and an external corona made primarily of free arms. To represent the scattering intensity of such particles, our model features two distinct regimes, taking place at different length scales. Therefore, following the standard assumptions adopted for deriving the form factors of colloidal star polymers and microgel particles~\cite{dozierColloidalNatureStar1991,clara-raholaStructuralPropertiesThermoresponsive2012}, we directly sum the corresponding scattering intensities and neglect any possible interference between them, as:
\begin{equation}
    P(q)= A_1P_{mgel}(q) + A_2P_{star}(q).
    \label{eq:model}
\end{equation}
Here, the first term, $P_{mgel}$, describes the overall structure of the particle, whereas the second one, $P_{star}$, accounts for the short-range correlations inside the star corona. The two contributions are weighted by the constants $A_1$ and $A_2$. 
\begin{figure}[b!]
    \centering
    \includegraphics[width=0.45\linewidth]{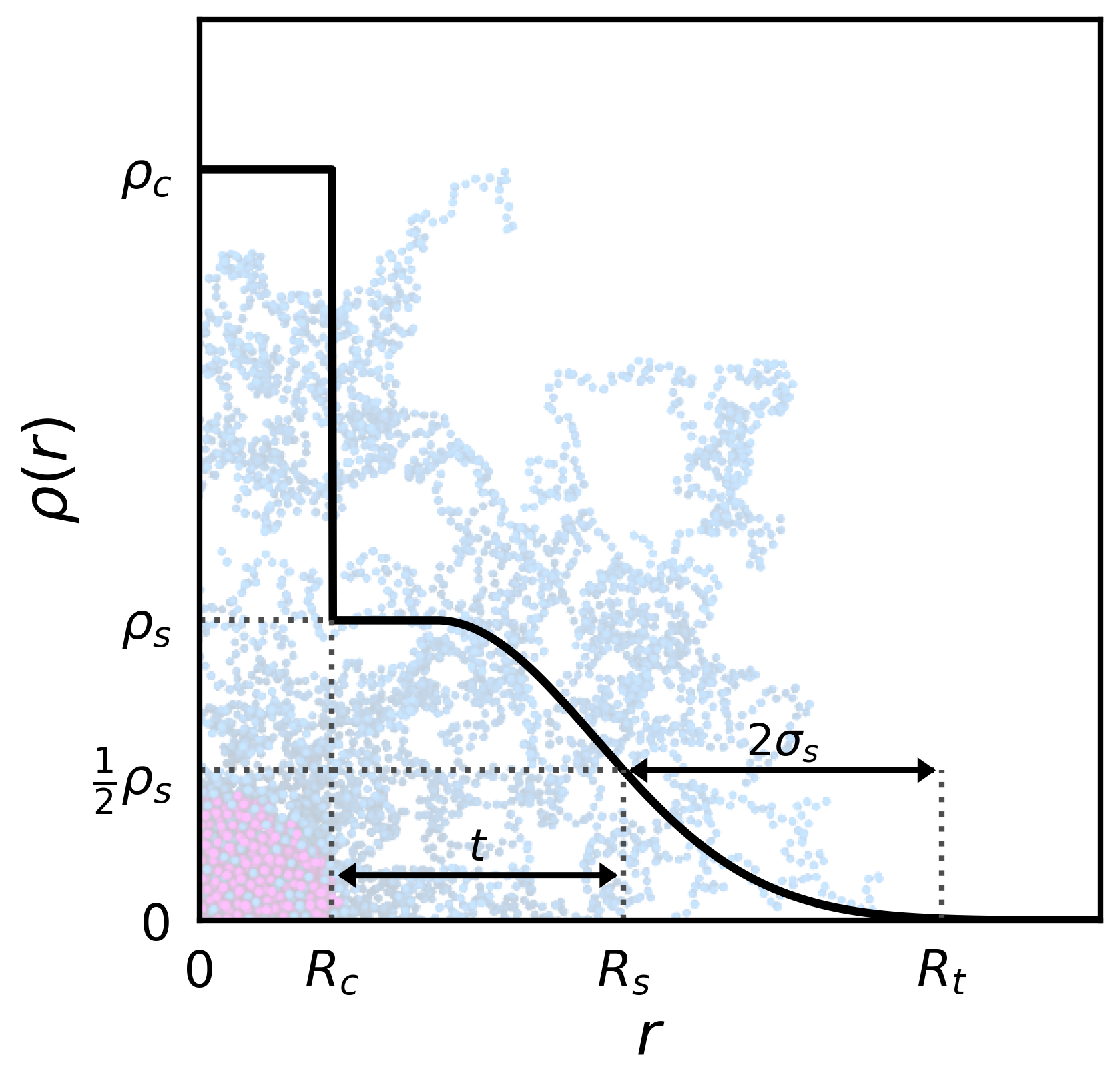}
    \caption{Sketch of the radial density profile of the ``core-fuzzy shell'' model for star-like microgels.}
    \label{fig:figura7}
\end{figure}

For $P_{mgel}(q)$, that is dominant at low $q$, we use a ``core-fuzzy shell'' model, consisting in a spherical core with radius $R_c$ and  scattering length density $\rho_c$, and a fuzzy shell with external radius $R_s$, fuzziness parameter $\sigma_s$ and  scattering length density $\rho_s$. This allows to account for the fuzzy edge of the star corona, in analogy to ref.~\citenum{stiegerSmallangleNeutronScattering2004}, and for its different contrast respect to the crosslinked core. The analytical expression of $P_{mgel}(q)$, sketched in Fig.\ref{fig:figura7}, is:
\begin{equation}
    P_{mgel}(q)= \frac{1}{V_{tot}}\,\left [ (\rho_c-\rho_s)A_s(q,R_c)+(\rho_s-\rho_0)A_s(q,R_s)\exp\left(-\frac{1}{2}\sigma_s^2q^2\right) \right ]^2\quad ,
    \label{eq:core-fuzzy-shell_model}
\end{equation}

where $\rho_0$ is the scattering length density of the solvent, $V_{tot}=\frac{4}{3}\pi R_t^3$ is total the volume of the particle ($R_t\approx R_s+2\sigma_s$~\cite{stiegerSmallangleNeutronScattering2004}), and $A_s(q,R)$ represents the scattering amplitude of a homogeneous sphere of radius $R$, written as:
\begin{equation}
    A_s(q,R) = 3V\,\frac{\sin(qR)-qR \cos(qR)}{(qR)^3}\quad ,
\end{equation}
where $V=\frac{4}{3}\pi R^3$. Within this model we thus define the thickness of the shell $t=R_s-R_c$. 
For $P_{star}(q)$, we use the expression provided by Dozier \textit{et al.}~\cite{dozierColloidalNatureStar1991}:
\begin{equation}
    P_{star}(q) = \frac{4\pi}{q\xi} \frac{\sin[\mu \arctan(q\xi)]}{(1+q^2\xi^2)^{\frac{\mu}{2}}} \Gamma(\frac{\mu}{2}) \quad ,
    \label{eq:star_model}
\end{equation}
where $\mu = (1 / \nu) - 1$, with $\nu$ the Flory solvency parameter, and $\xi$ is the maximum size of the polymer chain blobs, the spherical regions within which each arm of the star exhibits single-chain behavior~\cite{daoud1982star}, representing the characteristic length scale at which the granular structure becomes relevant. Hence, this term is dominant at large $q$. 
Since almost all of the scattering intensity comes from the largest blobs, namely the outermost ones \cite{dozierColloidalNatureStar1991}, there is no need to adjust this term, as the fact that the chains stem from the surface of the core rather than from the particle center has no significant effect.

It is worth noting that in the model for colloidal star polymers of Dozier \textit{et al.}~\cite{dozierColloidalNatureStar1991} the form factor is described by the sum of $P_{star}(q)$ and of a Guinier term $P_G(q)$, that only accounts for the gyration radius $R_g$ of the particle:
\begin{equation}
    P_G(q) \propto \exp\left( -\frac{1}{3}q^2R_g^2 \right) \quad,
    \label{eq:guinier}
\end{equation}
which has been replaced in the present analysis by the core-fuzzy shell term $P_{mgel}$.
However, when we fit the data with the simple star model in Fig.~S2(a) we refer to the original Dozier model, including the Guinier term.

\section*{Numerical Methods}

\subsection*{Design of Star Polymers}
We simulate star polymers with a central core surrounded by $f$ arms, each made on $N_f$ identical beads of diameter $\sigma$.
We adopt two strategies: in the first one, the core is composed of a single particle of radius $R_c$ and in the second one, which is more suitable to compare to experiments in the case of a large core at high EGDMA concentration, we make a core that is formed by several beads, again of radius $R_c$, to mimic the microgel situation, where the core is not a single particle. In this second approach, the core is designed by uniformly generating monomer positions within a sphere using a random sequential addition algorithm that maintains a minimum distance to prevent overlaps. A KDTree is then constructed to efficiently locate nearby monomers, forming bonds based on a cutoff distance. Each monomer is limited to a maximum of four bonds; any excess connections are pruned by removing the longest bond iteratively. 
Each arm is then anchored by selecting a core atom as the attachment point, with the constraint that each core atom can serve as a single attachment point.
The radial vector from the origin to the chosen core atom defines the chain’s orientation. The first monomer is placed at a fixed distance from the core, and subsequent monomers are sequentially positioned along the same radial direction at intervals equal to the bond length, avoiding particle overlaps.
Once the starting configuration is obtained, it is then run with the LAMMPS simulation software as described in the following also for microgels.

\subsection*{\textit{In silico} synthesis of Microgels}
To make PNIPAM-EGDMA microgels \textit{in silico} we exploit 
our previous assembly method developed for standard (PNIPAM-BIS) microgels~\cite{gnan2017silico,ninarello2019modeling}. In particular, we start using a binary mixture of $N=42000$ patchy particles of mass $m$ and diameter $\sigma$, respectively the units of mass and length, within a spherical cavity of radius $Z$. In particular, monomers are represented as divalent particles, while crosslinkers have a valence of four. The  molar ratio of crosslinkers $C=N_c/N$ used in experiments is the same as the experimental one, varying from 1\% to 10\%.

To reproduce the experimental fuzzy sphere structure of standard microgels, an additional force  is added on the crosslinkers, which mimics their faster reactivity with respect to NIPAM monomers. This radial force was optimized for BIS crosslinkers in Ref.~\citenum{ninarello2019modeling} to take the following form:
\begin{equation}
  \label{forces_first}
  \vec{f}_{d}=
  \begin{cases}
    -k r\hat{r}\ \ \        &\text{if}\ 0<r\leq D \\
    -g \hat{r}\ \ \     &\text{if}\ D <r<Z,
  \end{cases}
\end{equation}
where $\hat{r}$ is a unit vector pointing outward from the center of the cavity, 
 $D=Z/2$ is an intermediate length within which crosslinkers are mostly confined, and $k_{BIS}=4.5\times 10^{-5}\epsilon/\sigma^2$ and $g_{BIS}=0.008\epsilon/\sigma$ two phenomenological constants, where $\sigma$ and $\epsilon$ are units of length and energy, respectively.
The monomer number density $\rho_{mon}$ is constant to the value 
$\rho_{mon}=\frac{3}{4\pi}\frac{N}{Z^3}\simeq 0.08 \sigma^{-3}$
whose value was validated against experiments at different crosslinker concentrations~\cite{hazra2023structure}.

This methodology needs to be modified in order to take into account the much larger reactivity of EGDMA. In particular, we varied the parameters of radial force acting on the crosslinkers, while maintaining it still of the same form of Eq.~\ref{forces_first}. We find that to reproduce PNIPAM-EGDMA microgels we need to use
values of the constants in the force that are two orders of magnitude larger than those  employed for standard PNIPAM-BIS microgels, i.e.
\begin{eqnarray}
 k_{EGDMA}&=&4.5\times 10^{-3}\epsilon/\sigma^2;\\ 
 g_{EGDMA}&=&0.8\epsilon/\sigma.   
\end{eqnarray}
In addition, we also find that it is crucial to allow crosslinkers to bind among themselves as well as with NIPAM, something that could be neglected in the case of standard microgels, but is compulsory here in order to be able to assemble the microgels. Finally, we need to use a much more reduced size of the confining length $D$, to mimic the small size of the core made up of crosslinkers, which needs to increase with crosslinker concentration $C$ as $\sim C^{1/3}$. We thus determine $D=2\sigma$ and $D=4.3\sigma$ for $c=1\%$ and $c=10\%$, respectively. This recipe is thus in principle applicable to any microgel of this kind by simply rescaling $D$ with $C$ and $N$, both with a spherical growth law with exponent $(1/3)$. The monomer number density is always kept fixed to $\rho_{mon}.$
Once the force on the crosslinkers is set, assembly simulations are conducted at low temperature employing the oxDNA package~\cite{rovigatti2015comparison} with the additional swap protocol~\cite{sciortino2017three} to facilitate network formation as in previous works~\cite{gnan2017silico}. We wait until $>>99\%$ of the possible bonds is formed and then we retain the largest cluster only.
Importantly, we find that all input crosslinkers react and go to the core of the assembled structure, but there is quite a waste of the remaining monomers during the \textit{in silico} synthesis, so that we end up with the final number of beads in the microgel reducing to $\sim 30000$ and $\sim 35000$ for $c=1\%$ and $c=10\%$, respectively. 

\subsection*{Interaction potentials}
Once the star polymer is built or the microgel is assembled, the interactions between all particles (also called beads in the following), including monomers, crosslinkers and also the core of the stars are modeled with the bead-spring model~\cite{grest1986}, which is able to capture the behavior of polymeric particles in good solvent. In particular, all beads interact via a steric repulsion, modeled as a Weeks-Chandler-Anderson (WCA) potential%~\cite{weeks1971role}
\begin{equation}
V_{\rm WCA}(r)=
\begin{cases}
4\epsilon\left[\left(\frac{\sigma}{r}\right)^{12}-\left(\frac{\sigma}{r}\right)^{6}\right] + \epsilon & \text{if $r \le 2^{\frac{1}{6}}\sigma$}\\
0 & \text{otherwise,}
\end{cases}
\end{equation}
\noindent with $\epsilon$ setting the energy scale and $r$ the distance between two particles.  Bonded beads also experience the Finitely Extensible Nonlinear Elastic (FENE) potential
\begin{eqnarray}
V_{\rm FENE}(r)=
-\epsilon k_FR_0^2\ln\left[1-\left(\frac{r}{R_0\sigma}\right)^2\right]     \text{ if $r < R_0\sigma$,}
\end{eqnarray}
\noindent with $k_{\rm F}=15$ determining the stiffness of the bond and $R_{\rm 0} = 1.5$ the maximum bond distance. These equations are generalized by the standard mixed terms for unequal sizes when dealing with the core of the stars interacting with all other monomers.

In order to capture the affinity of the polymer to the solvent, we use the so-called solvophobic potential~\cite{soddemann2001}, commonly employed for PNIPAM-BIS microgels~\cite{gnan2017silico}:
\begin{equation}
  \label{va_force}
  V_\alpha(r)=
  \begin{cases}
    -\epsilon \alpha \ \ \        &\text{if}\ r\leq 2^{1/6}\sigma,\\
    \frac{1}{2}\alpha \epsilon  \big[ \cos(\gamma(r/\sigma)^2+\beta)-1 \big]  \ \ \     &\text{if}\ 2^{1/6}\sigma<r<R_0\sigma,\\
    0 \ \ \ &\text{if}\ r>R_0\sigma.
  \end{cases} 
\end{equation}
with $\gamma=\pi(2.25-2^{1/3})^{-1}$, with $\beta = 2\pi-2.25\gamma$. 
This effective attraction, implicitly modeling the solvent effect, is controlled by the solvophobic parameter $\alpha$, which plays the role of an effective temperature. For $\alpha=0$, the microgel is maximally swollen and no attraction in present, while, upon increasing $\alpha$, the microgel collapses. The Volume phase transition for this model occurs at $\alpha\sim 0.65$~\cite{ninarello2019modeling,gnan2017silico} and a linear mapping between $\alpha$ and the real temperature of experiments have been established for PNIPAM-BIS microgels in Refs.~\cite{ninarello2019modeling,hazra2023structure}, here also reported in the inset of Fig.~\ref{fig:figura6}.
We have further tested the variant in which EGDMA crosslinkers remain solvophilic at all temperatures, finding basically no difference in the results with respect to the case in which they are thermoresponsive. Therefore, for simplicity, we adopt the last modeling approach.

\subsection*{Simulations and calculated quantities}
Molecular dynamics simulations are performed in the NVT ensemble with a reduced constant temperature $T^*=k_BT/\epsilon=1$, with $k_B$ the Boltzmann constant and $T$ the temperature, that is controlled by a Nosé-Hoover thermostat. 
The simulated macromolecule, either a star or a microgel, is placed in the center of the simulation box with side $400\sigma$ and periodic boundary conditions. 
Equilibration runs are carried out for at least $1\times10^6 \delta t$, with $\delta t=0.002\tau$ and $\tau=\sqrt{m \sigma^2/\epsilon}$ the unit of time, followed by  production runs for at least $10\times10^6 \delta t$. All simulations are performed using the \textsc{lammps} package~\cite{thompson2022lammps}.

From the coordinates of all the beads, we then calculate the form factor $P(\mathbf{q})$ as:
\begin{equation}
P(\mathbf{q}) = \left\langle
\frac{1}{N_t} \sum_{i=1}^{N} \sum_{j=1}^{N}
\exp\!\bigl[i\,\mathbf{q}\cdot(\mathbf{r}_i - \mathbf{r}_j)\bigr]
\right\rangle,
\end{equation}
where the brackets denote an average over different configurations.
Similarly, we also compute the total density profile $\rho(r)$ as a function of the distance $r$ from the center of mass, as
\begin{equation}
    \label{eq:rho}
   \rho(r) = \langle \sum_{i=1}^{N} \delta (\left| \vec{r}-\vec{r}_i \right|)  \rangle.
\end{equation}
The density profiles of the crosslinkers are calculated considering the sum in Eq.~\ref{eq:rho} only up to the number of crosslinkers $N_c$.
The density profiles closer to star conditions should be described by a $r^{-4/3}$ dependence at intermediate distances, according to the Daoud-Cotton model in the swollen regime~\cite{likos2006soft}. However at large distances, there is always a more diffuse layer of polymers that is assumed to follow a Gaussian decay~\cite{mayer2007coarse}. Therefore, the density profiles are described according to the following functional form\cite{ruiz-francoMultiparticleCollisionDynamics2019}:
\begin{equation}
\label{eq:Rho_Star}
\rho^{star}\left(r\right)\sim A_{1}\,r^{-4/3}\,f_{b}\left(r,r_{b}\right)+ A_{2}\left[1-f_{b}\left(r,r_{b}\right)\right]\exp\left[-\left(\frac{r-r_{1}}{r_{2}} \right)^{2}\right],
\end{equation}
with $f_{b}\left(r,r_{b}\right)= \exp\left[-\left(\frac{r}{r_{b}}\right)^{4}\right]$ a bridge function between the two regimes.

Finally, the radius of gyration $R_g$, which gives a measure of the size of the particle is  calculated as,
\begin{equation}
R_{g} = \bigg\langle \left(\frac{1}{N_{m}}\sum_i^{N_{m}}(\vec{r}_i - \vec{r}_{\mathrm{cm}})^2\right)^{1/2} \bigg\rangle,
\end{equation} 
where $\vec{r_i}$ refers to the position of the $i$-th monomer and $\vec{r}_{\mathrm{cm}}$ to the center of mass of the star or of the microgel.

\begin{acknowledgement}
We thank D. Truzzolillo for useful discussions and C. Fedeli for preliminary simulations. We acknowledge the European Synchrotron Radiation Facility (ESRF) for provision of synchrotron radiation facilities  under proposal number SC-5593. EB, TP, ML and EZ acknowledge financial support by Progetto Co-MGELS funded by the European Union - NextGeneration EU under the National Recovery and Resilience Plan (PNRR) Mission 4 “Istruzione e Ricerca” - Component C2 - Investment 1.1 - "Fondo PRIN", Project code PRIN2022PAYLXW Sector PE11, CUP B53D23008890006. EB, FB, SS and EZ also acknowledge financial support from INAIL, project MicroMet (BRiC 2022, ID 16). 
\end{acknowledgement}

\bibliography{egdma}

\providecommand{\latin}[1]{#1}
\makeatletter
\providecommand{\doi}
  {\begingroup\let\do\@makeother\dospecials
  \catcode`\{=1 \catcode`\}=2 \doi@aux}
\providecommand{\doi@aux}[1]{\endgroup\texttt{#1}}
\makeatother
\providecommand*\mcitethebibliography{\thebibliography}
\csname @ifundefined\endcsname{endmcitethebibliography}  {\let\endmcitethebibliography\endthebibliography}{}
\begin{mcitethebibliography}{59}
\providecommand*\natexlab[1]{#1}
\providecommand*\mciteSetBstSublistMode[1]{}
\providecommand*\mciteSetBstMaxWidthForm[2]{}
\providecommand*\mciteBstWouldAddEndPuncttrue
  {\def\EndOfBibitem{\unskip.}}
\providecommand*\mciteBstWouldAddEndPunctfalse
  {\let\EndOfBibitem\relax}
\providecommand*\mciteSetBstMidEndSepPunct[3]{}
\providecommand*\mciteSetBstSublistLabelBeginEnd[3]{}
\providecommand*\EndOfBibitem{}
\mciteSetBstSublistMode{f}
\mciteSetBstMaxWidthForm{subitem}{(\alph{mcitesubitemcount})}
\mciteSetBstSublistLabelBeginEnd
  {\mcitemaxwidthsubitemform\space}
  {\relax}
  {\relax}

\bibitem[Likos \latin{et~al.}(1998)Likos, L{\"o}wen, Watzlawek, Abbas, Jucknischke, Allgaier, and Richter]{likos1998star}
Likos,~C.~N.; L{\"o}wen,~H.; Watzlawek,~M.; Abbas,~B.; Jucknischke,~O.; Allgaier,~J.; Richter,~D. Star polymers viewed as ultrasoft colloidal particles. \emph{Phys. Rev. Lett.} \textbf{1998}, \emph{80}, 4450\relax
\mciteBstWouldAddEndPuncttrue
\mciteSetBstMidEndSepPunct{\mcitedefaultmidpunct}
{\mcitedefaultendpunct}{\mcitedefaultseppunct}\relax
\EndOfBibitem
\bibitem[Vlassopoulos(2004)]{vlassopoulos2004colloidal}
Vlassopoulos,~D. Colloidal star polymers: Models for studying dynamically arrested states in soft matter. \emph{Journal of Polymer Science Part B: Polymer Physics} \textbf{2004}, \emph{42}, 2931--2941\relax
\mciteBstWouldAddEndPuncttrue
\mciteSetBstMidEndSepPunct{\mcitedefaultmidpunct}
{\mcitedefaultendpunct}{\mcitedefaultseppunct}\relax
\EndOfBibitem
\bibitem[Ren \latin{et~al.}(2016)Ren, McKenzie, Fu, Wong, Xu, An, Shanmugam, Davis, Boyer, and Qiao]{renStarPolymers2016}
Ren,~J.~M.; McKenzie,~T.~G.; Fu,~Q.; Wong,~E. H.~H.; Xu,~J.; An,~Z.; Shanmugam,~S.; Davis,~T.~P.; Boyer,~C.; Qiao,~G.~G. Star {{Polymers}}. \emph{Chemical Reviews} \textbf{2016}, \emph{116}, 6743--6836\relax
\mciteBstWouldAddEndPuncttrue
\mciteSetBstMidEndSepPunct{\mcitedefaultmidpunct}
{\mcitedefaultendpunct}{\mcitedefaultseppunct}\relax
\EndOfBibitem
\bibitem[Wu \latin{et~al.}(2015)Wu, Wang, and Li]{wuStarPolymersAdvances2015}
Wu,~W.; Wang,~W.; Li,~J. Star Polymers: {{Advances}} in Biomedical Applications. \emph{Progress in Polymer Science} \textbf{2015}, \emph{46}, 55--85\relax
\mciteBstWouldAddEndPuncttrue
\mciteSetBstMidEndSepPunct{\mcitedefaultmidpunct}
{\mcitedefaultendpunct}{\mcitedefaultseppunct}\relax
\EndOfBibitem
\bibitem[Likos(2006)]{likos2006soft}
Likos,~C.~N. Soft matter with soft particles. \emph{Soft Matter} \textbf{2006}, \emph{2}, 478--498\relax
\mciteBstWouldAddEndPuncttrue
\mciteSetBstMidEndSepPunct{\mcitedefaultmidpunct}
{\mcitedefaultendpunct}{\mcitedefaultseppunct}\relax
\EndOfBibitem
\bibitem[Roovers \latin{et~al.}(1989)Roovers, Toporowski, and Martin]{roovers1989synthesis}
Roovers,~J.; Toporowski,~P.; Martin,~J. Synthesis and characterization of multiarm star polybutadienes. \emph{Macromolecules} \textbf{1989}, \emph{22}, 1897--1903\relax
\mciteBstWouldAddEndPuncttrue
\mciteSetBstMidEndSepPunct{\mcitedefaultmidpunct}
{\mcitedefaultendpunct}{\mcitedefaultseppunct}\relax
\EndOfBibitem
\bibitem[Connal \latin{et~al.}(2008)Connal, Li, Quinn, Tjipto, Caruso, and Qiao]{connal2008}
Connal,~L.~A.; Li,~Q.; Quinn,~J.~F.; Tjipto,~E.; Caruso,~F.; Qiao,~G.~G. pH-Responsive Poly(acrylic acid) Core Cross-Linked Star Polymers: Morphology Transitions in Solution and Multilayer Thin Films. \emph{Macromolecules} \textbf{2008}, \emph{41}, 2620--2626\relax
\mciteBstWouldAddEndPuncttrue
\mciteSetBstMidEndSepPunct{\mcitedefaultmidpunct}
{\mcitedefaultendpunct}{\mcitedefaultseppunct}\relax
\EndOfBibitem
\bibitem[Lang \latin{et~al.}(2017)Lang, Lenart, Sun, Hammouda, and Hore]{lang2017}
Lang,~X.; Lenart,~W.~R.; Sun,~J. E.~P.; Hammouda,~B.; Hore,~M. J.~A. Interaction and Conformation of Aqueous Poly(N-isopropylacrylamide) (PNIPAM) Star Polymers below the LCST. \emph{Macromolecules} \textbf{2017}, \emph{50}, 2145--2154\relax
\mciteBstWouldAddEndPuncttrue
\mciteSetBstMidEndSepPunct{\mcitedefaultmidpunct}
{\mcitedefaultendpunct}{\mcitedefaultseppunct}\relax
\EndOfBibitem
\bibitem[Cao \latin{et~al.}(2018)Cao, Han, Duan, and Zhang]{Cao2018}
Cao,~M.; Han,~G.; Duan,~W.; Zhang,~W. Synthesis of multi-arm star thermo-responsive polymers and topology effects on phase transition. \emph{Polym. Chem.} \textbf{2018}, \emph{9}, 2625--2633\relax
\mciteBstWouldAddEndPuncttrue
\mciteSetBstMidEndSepPunct{\mcitedefaultmidpunct}
{\mcitedefaultendpunct}{\mcitedefaultseppunct}\relax
\EndOfBibitem
\bibitem[Okaya \latin{et~al.}(2020)Okaya, Jochi, Seki, Satoh, Kamigaito, Hoshino, Nakatani, Fujinami, Takata, and Takeoka]{okaya2020}
Okaya,~Y.; Jochi,~Y.; Seki,~T.; Satoh,~K.; Kamigaito,~M.; Hoshino,~T.; Nakatani,~T.; Fujinami,~S.; Takata,~M.; Takeoka,~Y. Precise Synthesis of a Homogeneous Thermoresponsive Polymer Network Composed of Four-Branched Star Polymers with a Narrow Molecular Weight Distribution. \emph{Macromolecules} \textbf{2020}, \emph{53}, 374--386\relax
\mciteBstWouldAddEndPuncttrue
\mciteSetBstMidEndSepPunct{\mcitedefaultmidpunct}
{\mcitedefaultendpunct}{\mcitedefaultseppunct}\relax
\EndOfBibitem
\bibitem[Kubota \latin{et~al.}(1990)Kubota, Fujishige, and Ando]{kubotaSinglechainTransitionPolyNisopropylacrylamide1990}
Kubota,~{\relax Kenji}.; Fujishige,~{\relax Shouei}.; Ando,~{\relax Isao}. Single-Chain Transition of Poly({{N-isopropylacrylamide}}) in Water. \emph{The Journal of Physical Chemistry} \textbf{1990}, \emph{94}, 5154--5158\relax
\mciteBstWouldAddEndPuncttrue
\mciteSetBstMidEndSepPunct{\mcitedefaultmidpunct}
{\mcitedefaultendpunct}{\mcitedefaultseppunct}\relax
\EndOfBibitem
\bibitem[Fernandez-Nieves \latin{et~al.}(2011)Fernandez-Nieves, Wyss, Mattsson, and Weitz]{fernandez2011microgel}
Fernandez-Nieves,~A.; Wyss,~H.; Mattsson,~J.; Weitz,~D.~A. \emph{Microgel suspensions: fundamentals and applications}; John Wiley \& Sons, 2011\relax
\mciteBstWouldAddEndPuncttrue
\mciteSetBstMidEndSepPunct{\mcitedefaultmidpunct}
{\mcitedefaultendpunct}{\mcitedefaultseppunct}\relax
\EndOfBibitem
\bibitem[Pelton(2000)]{pelton2000temperature}
Pelton,~R. Temperature-sensitive aqueous microgels. \emph{Advances in colloid and interface science} \textbf{2000}, \emph{85}, 1--33\relax
\mciteBstWouldAddEndPuncttrue
\mciteSetBstMidEndSepPunct{\mcitedefaultmidpunct}
{\mcitedefaultendpunct}{\mcitedefaultseppunct}\relax
\EndOfBibitem
\bibitem[Stieger \latin{et~al.}(2004)Stieger, Richtering, Pedersen, and Lindner]{stiegerSmallangleNeutronScattering2004}
Stieger,~M.; Richtering,~W.; Pedersen,~J.~S.; Lindner,~P. Small-Angle Neutron Scattering Study of Structural Changes in Temperature Sensitive Microgel Colloids. \emph{The Journal of Chemical Physics} \textbf{2004}, \emph{120}, 6197--6206\relax
\mciteBstWouldAddEndPuncttrue
\mciteSetBstMidEndSepPunct{\mcitedefaultmidpunct}
{\mcitedefaultendpunct}{\mcitedefaultseppunct}\relax
\EndOfBibitem
\bibitem[Hertle and Hellweg(2013)Hertle, and Hellweg]{hertleThermoresponsiveCopolymerMicrogels2013}
Hertle,~Y.; Hellweg,~T. Thermoresponsive Copolymer Microgels. \emph{Journal of Materials Chemistry B} \textbf{2013}, \emph{1}, 5874\relax
\mciteBstWouldAddEndPuncttrue
\mciteSetBstMidEndSepPunct{\mcitedefaultmidpunct}
{\mcitedefaultendpunct}{\mcitedefaultseppunct}\relax
\EndOfBibitem
\bibitem[Plamper and Richtering(2017)Plamper, and Richtering]{plamperFunctionalMicrogelsMicrogel2017b}
Plamper,~F.~A.; Richtering,~W. Functional {{Microgels}} and {{Microgel Systems}}. \emph{Accounts of Chemical Research} \textbf{2017}, \emph{50}, 131--140\relax
\mciteBstWouldAddEndPuncttrue
\mciteSetBstMidEndSepPunct{\mcitedefaultmidpunct}
{\mcitedefaultendpunct}{\mcitedefaultseppunct}\relax
\EndOfBibitem
\bibitem[Karg and Hellweg(2009)Karg, and Hellweg]{karg2009smart}
Karg,~M.; Hellweg,~T. Smart inorganic/organic hybrid microgels: Synthesis and characterisation. \emph{Journal of Materials Chemistry} \textbf{2009}, \emph{19}, 8714--8727\relax
\mciteBstWouldAddEndPuncttrue
\mciteSetBstMidEndSepPunct{\mcitedefaultmidpunct}
{\mcitedefaultendpunct}{\mcitedefaultseppunct}\relax
\EndOfBibitem
\bibitem[Gury \latin{et~al.}(2025)Gury, Gauthier, Suau, Vlassopoulos, and Cloitre]{gury2025}
Gury,~L.; Gauthier,~M.; Suau,~J.-M.; Vlassopoulos,~D.; Cloitre,~M. Internal Microstructure Dictates Yielding and Flow of Jammed Suspensions and Emulsions. \emph{ACS Nano} \textbf{2025}, \relax
\mciteBstWouldAddEndPunctfalse
\mciteSetBstMidEndSepPunct{\mcitedefaultmidpunct}
{}{\mcitedefaultseppunct}\relax
\EndOfBibitem
\bibitem[Mueller \latin{et~al.}(2018)Mueller, Alsop, Scotti, Bleuel, Rheinst\"adter, Richtering, and Hoare]{mueller2018dynamically}
Mueller,~E.; Alsop,~R.~J.; Scotti,~A.; Bleuel,~M.; Rheinst\"adter,~M.~C.; Richtering,~W.; Hoare,~T. Dynamically cross-linked self-assembled thermoresponsive microgels with homogeneous internal structures. \emph{Langmuir} \textbf{2018}, \emph{34}, 1601--1612\relax
\mciteBstWouldAddEndPuncttrue
\mciteSetBstMidEndSepPunct{\mcitedefaultmidpunct}
{\mcitedefaultendpunct}{\mcitedefaultseppunct}\relax
\EndOfBibitem
\bibitem[Kyrey \latin{et~al.}(2019)Kyrey, Witte, Feoktystov, Pipich, Wu, Pasini, Radulescu, Witt, Kruteva, von Klitzing, \latin{et~al.} others]{kyrey2019inner}
Kyrey,~T.; Witte,~J.; Feoktystov,~A.; Pipich,~V.; Wu,~B.; Pasini,~S.; Radulescu,~A.; Witt,~M.~U.; Kruteva,~M.; von Klitzing,~R.; others Inner structure and dynamics of microgels with low and medium crosslinker content prepared via surfactant-free precipitation polymerization and continuous monomer feeding approach. \emph{Soft Matter} \textbf{2019}, \emph{15}, 6536--6546\relax
\mciteBstWouldAddEndPuncttrue
\mciteSetBstMidEndSepPunct{\mcitedefaultmidpunct}
{\mcitedefaultendpunct}{\mcitedefaultseppunct}\relax
\EndOfBibitem
\bibitem[Still \latin{et~al.}(2013)Still, Chen, Alsayed, Aptowicz, and Yodh]{stillSynthesisMicrometersizePolyNisopropylacrylamide2013}
Still,~T.; Chen,~K.; Alsayed,~A.~M.; Aptowicz,~K.~B.; Yodh,~A. Synthesis of Micrometer-Size Poly({{N-isopropylacrylamide}}) Microgel Particles with Homogeneous Crosslinker Density and Diameter Control. \emph{Journal of Colloid and Interface Science} \textbf{2013}, \emph{405}, 96--102\relax
\mciteBstWouldAddEndPuncttrue
\mciteSetBstMidEndSepPunct{\mcitedefaultmidpunct}
{\mcitedefaultendpunct}{\mcitedefaultseppunct}\relax
\EndOfBibitem
\bibitem[Gao and Frisken(2003)Gao, and Frisken]{gao2003cross}
Gao,~J.; Frisken,~B.~J. Cross-linker-free N-isopropylacrylamide gel nanospheres. \emph{Langmuir} \textbf{2003}, \emph{19}, 5212--5216\relax
\mciteBstWouldAddEndPuncttrue
\mciteSetBstMidEndSepPunct{\mcitedefaultmidpunct}
{\mcitedefaultendpunct}{\mcitedefaultseppunct}\relax
\EndOfBibitem
\bibitem[Bachman \latin{et~al.}(2015)Bachman, Brown, Clarke, Dhada, Douglas, Hansen, Herman, Hyatt, Kodlekere, Meng, \latin{et~al.} others]{bachman2015ultrasoft}
Bachman,~H.; Brown,~A.~C.; Clarke,~K.~C.; Dhada,~K.~S.; Douglas,~A.; Hansen,~C.~E.; Herman,~E.; Hyatt,~J.~S.; Kodlekere,~P.; Meng,~Z.; others Ultrasoft, highly deformable microgels. \emph{Soft matter} \textbf{2015}, \emph{11}, 2018--2028\relax
\mciteBstWouldAddEndPuncttrue
\mciteSetBstMidEndSepPunct{\mcitedefaultmidpunct}
{\mcitedefaultendpunct}{\mcitedefaultseppunct}\relax
\EndOfBibitem
\bibitem[Kratz \latin{et~al.}(2002)Kratz, Lapp, Eimer, and Hellweg]{kratzVolumeTransitionStructure2002}
Kratz,~K.; Lapp,~A.; Eimer,~W.; Hellweg,~T. Volume Transition and Structure of Triethyleneglycol Dimethacrylate, Ethylenglykol Dimethacrylate, and {{N}},{{N}}{$\prime$}-Methylene Bis-Acrylamide Cross-Linked Poly({{N-isopropyl}} Acrylamide) Microgels: A Small Angle Neutron and Dynamic Light Scattering Study. \emph{Colloids and Surfaces A: Physicochemical and Engineering Aspects} \textbf{2002}, \emph{197}, 55--67\relax
\mciteBstWouldAddEndPuncttrue
\mciteSetBstMidEndSepPunct{\mcitedefaultmidpunct}
{\mcitedefaultendpunct}{\mcitedefaultseppunct}\relax
\EndOfBibitem
\bibitem[Gawlitza \latin{et~al.}(2014)Gawlitza, Radulescu, von Klitzing, and Wellert]{gawlitza2014structure}
Gawlitza,~K.; Radulescu,~A.; von Klitzing,~R.; Wellert,~S. On the structure of biocompatible, thermoresponsive poly (ethylene glycol) microgels. \emph{Polymer} \textbf{2014}, \emph{55}, 6717--6724\relax
\mciteBstWouldAddEndPuncttrue
\mciteSetBstMidEndSepPunct{\mcitedefaultmidpunct}
{\mcitedefaultendpunct}{\mcitedefaultseppunct}\relax
\EndOfBibitem
\bibitem[Aguirre \latin{et~al.}(2019)Aguirre, Deniau, Br{\^u}let, Chougrani, Alard, and Billon]{aguirre2019versatile}
Aguirre,~G.; Deniau,~E.; Br{\^u}let,~A.; Chougrani,~K.; Alard,~V.; Billon,~L. Versatile oligo (ethylene glycol)-based biocompatible microgels for loading/release of active bio (macro) molecules. \emph{Colloids and Surfaces B: Biointerfaces} \textbf{2019}, \emph{175}, 445--453\relax
\mciteBstWouldAddEndPuncttrue
\mciteSetBstMidEndSepPunct{\mcitedefaultmidpunct}
{\mcitedefaultendpunct}{\mcitedefaultseppunct}\relax
\EndOfBibitem
\bibitem[Rivas-Barbosa \latin{et~al.}(2022)Rivas-Barbosa, Ruiz-Franco, Lara-Pe\~na, Cardellini, Licea-Claverie, Camerin, Zaccarelli, and Laurati]{rivas2022link}
Rivas-Barbosa,~R.; Ruiz-Franco,~J.; Lara-Pe\~na,~M.~A.; Cardellini,~J.; Licea-Claverie,~A.; Camerin,~F.; Zaccarelli,~E.; Laurati,~M. Link between morphology, structure, and interactions of composite microgels. \emph{Macromolecules} \textbf{2022}, \emph{55}, 1834--1843\relax
\mciteBstWouldAddEndPuncttrue
\mciteSetBstMidEndSepPunct{\mcitedefaultmidpunct}
{\mcitedefaultendpunct}{\mcitedefaultseppunct}\relax
\EndOfBibitem
\bibitem[Del~Monte \latin{et~al.}(2021)Del~Monte, Truzzolillo, Camerin, Ninarello, Chauveau, Tavagnacco, Gnan, Rovigatti, Sennato, and Zaccarelli]{delmonteTwostepDeswellingVolume2021a}
Del~Monte,~G.; Truzzolillo,~D.; Camerin,~F.; Ninarello,~A.; Chauveau,~E.; Tavagnacco,~L.; Gnan,~N.; Rovigatti,~L.; Sennato,~S.; Zaccarelli,~E. Two-Step Deswelling in the {{Volume Phase Transition}} of Thermoresponsive Microgels. \emph{Proceedings of the National Academy of Sciences} \textbf{2021}, \emph{118}, e2109560118\relax
\mciteBstWouldAddEndPuncttrue
\mciteSetBstMidEndSepPunct{\mcitedefaultmidpunct}
{\mcitedefaultendpunct}{\mcitedefaultseppunct}\relax
\EndOfBibitem
\bibitem[Elancheliyan \latin{et~al.}(2022)Elancheliyan, Del~Monte, Chauveau, Sennato, Zaccarelli, and Truzzolillo]{elancheliyanRoleChargeContent2022a}
Elancheliyan,~R.; Del~Monte,~G.; Chauveau,~E.; Sennato,~S.; Zaccarelli,~E.; Truzzolillo,~D. Role of {{Charge Content}} in the {{Two-Step Deswelling}} of {{Poly}}( {{{\emph{N}}}} -Isopropylacrylamide)-{{Based Microgels}}. \emph{Macromolecules} \textbf{2022}, \emph{55}, 7526--7539\relax
\mciteBstWouldAddEndPuncttrue
\mciteSetBstMidEndSepPunct{\mcitedefaultmidpunct}
{\mcitedefaultendpunct}{\mcitedefaultseppunct}\relax
\EndOfBibitem
\bibitem[Dozier \latin{et~al.}(1991)Dozier, Huang, and Fetters]{dozierColloidalNatureStar1991}
Dozier,~W.~D.; Huang,~J.~S.; Fetters,~L.~J. Colloidal Nature of Star Polymer Dilute and Semidilute Solutions. \emph{Macromolecules} \textbf{1991}, \emph{24}, 2810--2814\relax
\mciteBstWouldAddEndPuncttrue
\mciteSetBstMidEndSepPunct{\mcitedefaultmidpunct}
{\mcitedefaultendpunct}{\mcitedefaultseppunct}\relax
\EndOfBibitem
\bibitem[Ninarello \latin{et~al.}(2019)Ninarello, Crassous, Paloli, Camerin, Gnan, Rovigatti, Schurtenberger, and Zaccarelli]{ninarello2019modeling}
Ninarello,~A.; Crassous,~J.~J.; Paloli,~D.; Camerin,~F.; Gnan,~N.; Rovigatti,~L.; Schurtenberger,~P.; Zaccarelli,~E. Modeling microgels with a controlled structure across the volume phase transition. \emph{Macromolecules} \textbf{2019}, \emph{52}, 7584--7592\relax
\mciteBstWouldAddEndPuncttrue
\mciteSetBstMidEndSepPunct{\mcitedefaultmidpunct}
{\mcitedefaultendpunct}{\mcitedefaultseppunct}\relax
\EndOfBibitem
\bibitem[Hazra \latin{et~al.}(2023)Hazra, Ninarello, Scotti, Houston, Mota-Santiago, Zaccarelli, and Crassous]{hazra2023structure}
Hazra,~N.; Ninarello,~A.; Scotti,~A.; Houston,~J.~E.; Mota-Santiago,~P.; Zaccarelli,~E.; Crassous,~J.~J. Structure of Responsive Microgels down to Ultralow Cross-Linkings. \emph{Macromolecules} \textbf{2023}, \emph{57}, 339--355\relax
\mciteBstWouldAddEndPuncttrue
\mciteSetBstMidEndSepPunct{\mcitedefaultmidpunct}
{\mcitedefaultendpunct}{\mcitedefaultseppunct}\relax
\EndOfBibitem
\bibitem[Hildebrandt \latin{et~al.}(2023)Hildebrandt, Thuy, Kippenberger, Wigger, Houston, Scotti, and Karg]{hildebrandt2023fluid}
Hildebrandt,~M.; Thuy,~D.~P.; Kippenberger,~J.; Wigger,~T.~L.; Houston,~J.~E.; Scotti,~A.; Karg,~M. Fluid--solid transitions in photonic crystals of soft, thermoresponsive microgels. \emph{Soft Matter} \textbf{2023}, \emph{19}, 7122--7135\relax
\mciteBstWouldAddEndPuncttrue
\mciteSetBstMidEndSepPunct{\mcitedefaultmidpunct}
{\mcitedefaultendpunct}{\mcitedefaultseppunct}\relax
\EndOfBibitem
\bibitem[Serrano-Medina \latin{et~al.}(2012)Serrano-Medina, Cornejo-Bravo, and Licea-Claver{\'\i}e]{serrano2012synthesis}
Serrano-Medina,~A.; Cornejo-Bravo,~J.; Licea-Claver{\'\i}e,~A. Synthesis of pH and temperature sensitive, core--shell nano/microgels, by one pot, soap-free emulsion polymerization. \emph{Journal of colloid and interface science} \textbf{2012}, \emph{369}, 82--90\relax
\mciteBstWouldAddEndPuncttrue
\mciteSetBstMidEndSepPunct{\mcitedefaultmidpunct}
{\mcitedefaultendpunct}{\mcitedefaultseppunct}\relax
\EndOfBibitem
\bibitem[{Clara-Rahola} \latin{et~al.}(2012){Clara-Rahola}, {Fernandez-Nieves}, {Sierra-Martin}, South, Lyon, Kohlbrecher, and Fernandez~Barbero]{clara-raholaStructuralPropertiesThermoresponsive2012}
{Clara-Rahola},~J.; {Fernandez-Nieves},~A.; {Sierra-Martin},~B.; South,~A.~B.; Lyon,~L.~A.; Kohlbrecher,~J.; Fernandez~Barbero,~A. Structural Properties of Thermoresponsive Poly( {{{\emph{N}}}} -Isopropylacrylamide)-Poly(Ethyleneglycol) Microgels. \emph{The Journal of Chemical Physics} \textbf{2012}, \emph{136}, 214903\relax
\mciteBstWouldAddEndPuncttrue
\mciteSetBstMidEndSepPunct{\mcitedefaultmidpunct}
{\mcitedefaultendpunct}{\mcitedefaultseppunct}\relax
\EndOfBibitem
\bibitem[Sayed \latin{et~al.}(2019)Sayed, Lorthioir, Perrin, and Sanson]{sayed2019pegylated}
Sayed,~J.~E.; Lorthioir,~C.; Perrin,~P.; Sanson,~N. PEGylated NiPAM microgels: synthesis, characterization and colloidal stability. \emph{Soft Matter} \textbf{2019}, \emph{15}, 963--972\relax
\mciteBstWouldAddEndPuncttrue
\mciteSetBstMidEndSepPunct{\mcitedefaultmidpunct}
{\mcitedefaultendpunct}{\mcitedefaultseppunct}\relax
\EndOfBibitem
\bibitem[{Ruiz-Franco} \latin{et~al.}(2023){Ruiz-Franco}, {Rivas-Barbosa}, {Lara-Pe{\~n}a}, {Villanueva-Valencia}, {Licea-Claverie}, Zaccarelli, and Laurati]{ruiz-francoConcentrationTemperatureDependent2023c}
{Ruiz-Franco},~J.; {Rivas-Barbosa},~R.; {Lara-Pe{\~n}a},~M.~A.; {Villanueva-Valencia},~J.~R.; {Licea-Claverie},~A.; Zaccarelli,~E.; Laurati,~M. Concentration and Temperature Dependent Interactions and State Diagram of Dispersions of Copolymer Microgels. \emph{Soft Matter} \textbf{2023}, \emph{19}, 3614--3628\relax
\mciteBstWouldAddEndPuncttrue
\mciteSetBstMidEndSepPunct{\mcitedefaultmidpunct}
{\mcitedefaultendpunct}{\mcitedefaultseppunct}\relax
\EndOfBibitem
\bibitem[Vlassopoulos and Cloitre(2014)Vlassopoulos, and Cloitre]{vlassopoulos2014tunable}
Vlassopoulos,~D.; Cloitre,~M. Tunable rheology of dense soft deformable colloids. \emph{Current opinion in colloid \& interface science} \textbf{2014}, \emph{19}, 561--574\relax
\mciteBstWouldAddEndPuncttrue
\mciteSetBstMidEndSepPunct{\mcitedefaultmidpunct}
{\mcitedefaultendpunct}{\mcitedefaultseppunct}\relax
\EndOfBibitem
\bibitem[Van Der~Scheer \latin{et~al.}(2017)Van Der~Scheer, Van De~Laar, Van Der~Gucht, Vlassopoulos, and Sprakel]{van2017fragility}
Van Der~Scheer,~P.; Van De~Laar,~T.; Van Der~Gucht,~J.; Vlassopoulos,~D.; Sprakel,~J. Fragility and strength in nanoparticle glasses. \emph{ACS nano} \textbf{2017}, \emph{11}, 6755--6763\relax
\mciteBstWouldAddEndPuncttrue
\mciteSetBstMidEndSepPunct{\mcitedefaultmidpunct}
{\mcitedefaultendpunct}{\mcitedefaultseppunct}\relax
\EndOfBibitem
\bibitem[Gury \latin{et~al.}(2021)Gury, Kamble, Parisi, Zhang, Lee, Abdullah, Matyjaszewski, Bockstaller, Vlassopoulos, and Fytas]{gury2021internal}
Gury,~L.; Kamble,~S.; Parisi,~D.; Zhang,~J.; Lee,~J.; Abdullah,~A.; Matyjaszewski,~K.; Bockstaller,~M.~R.; Vlassopoulos,~D.; Fytas,~G. Internal Microstructure Dictates Interactions of Polymer-grafted Nanoparticles in Solution. \emph{Macromolecules} \textbf{2021}, \emph{54}, 7234--7243\relax
\mciteBstWouldAddEndPuncttrue
\mciteSetBstMidEndSepPunct{\mcitedefaultmidpunct}
{\mcitedefaultendpunct}{\mcitedefaultseppunct}\relax
\EndOfBibitem
\bibitem[Scotti \latin{et~al.}(2022)Scotti, Schulte, Lopez, Crassous, Bochenek, and Richtering]{scotti2022softness}
Scotti,~A.; Schulte,~M.~F.; Lopez,~C.~G.; Crassous,~J.~J.; Bochenek,~S.; Richtering,~W. How softness matters in soft nanogels and nanogel assemblies. \emph{Chemical reviews} \textbf{2022}, \emph{122}, 11675--11700\relax
\mciteBstWouldAddEndPuncttrue
\mciteSetBstMidEndSepPunct{\mcitedefaultmidpunct}
{\mcitedefaultendpunct}{\mcitedefaultseppunct}\relax
\EndOfBibitem
\bibitem[Camerin \latin{et~al.}(2019)Camerin, Fern\'andez-Rodr\'iguez, Rovigatti, Antonopoulou, Gnan, Ninarello, Isa, and Zaccarelli]{camerin2019microgels}
Camerin,~F.; Fern\'andez-Rodr\'iguez,~M.~{\'A}.; Rovigatti,~L.; Antonopoulou,~M.-N.; Gnan,~N.; Ninarello,~A.; Isa,~L.; Zaccarelli,~E. Microgels adsorbed at liquid--liquid interfaces: A joint numerical and experimental study. \emph{ACS nano} \textbf{2019}, \emph{13}, 4548--4559\relax
\mciteBstWouldAddEndPuncttrue
\mciteSetBstMidEndSepPunct{\mcitedefaultmidpunct}
{\mcitedefaultendpunct}{\mcitedefaultseppunct}\relax
\EndOfBibitem
\bibitem[Menath \latin{et~al.}(2021)Menath, Eatson, Brilmayer, Andrieu-Brunsen, Buzza, and Vogel]{menath2021defined}
Menath,~J.; Eatson,~J.; Brilmayer,~R.; Andrieu-Brunsen,~A.; Buzza,~D. M.~A.; Vogel,~N. Defined core--shell particles as the key to complex interfacial self-assembly. \emph{Proceedings of the National Academy of Sciences} \textbf{2021}, \emph{118}, e2113394118\relax
\mciteBstWouldAddEndPuncttrue
\mciteSetBstMidEndSepPunct{\mcitedefaultmidpunct}
{\mcitedefaultendpunct}{\mcitedefaultseppunct}\relax
\EndOfBibitem
\bibitem[Nikolov \latin{et~al.}(2020)Nikolov, Fernandez-Nieves, and Alexeev]{nikolov2020behavior}
Nikolov,~S.~V.; Fernandez-Nieves,~A.; Alexeev,~A. Behavior and mechanics of dense microgel suspensions. \emph{Proceedings of the National Academy of Sciences} \textbf{2020}, \emph{117}, 27096--27103\relax
\mciteBstWouldAddEndPuncttrue
\mciteSetBstMidEndSepPunct{\mcitedefaultmidpunct}
{\mcitedefaultendpunct}{\mcitedefaultseppunct}\relax
\EndOfBibitem
\bibitem[Del~Monte and Zaccarelli(2024)Del~Monte, and Zaccarelli]{del2024numerical}
Del~Monte,~G.; Zaccarelli,~E. Numerical study of neutral and charged microgel suspensions: From single-particle to collective behavior. \emph{Physical Review X} \textbf{2024}, \emph{14}, 041067\relax
\mciteBstWouldAddEndPuncttrue
\mciteSetBstMidEndSepPunct{\mcitedefaultmidpunct}
{\mcitedefaultendpunct}{\mcitedefaultseppunct}\relax
\EndOfBibitem
\bibitem[Koppel(1972)]{koppel1972analysis}
Koppel,~D.~E. Analysis of macromolecular polydispersity in intensity correlation spectroscopy: the method of cumulants. \emph{The Journal of Chemical Physics} \textbf{1972}, \emph{57}, 4814--4820\relax
\mciteBstWouldAddEndPuncttrue
\mciteSetBstMidEndSepPunct{\mcitedefaultmidpunct}
{\mcitedefaultendpunct}{\mcitedefaultseppunct}\relax
\EndOfBibitem
\bibitem[Narayanan \latin{et~al.}(2022)Narayanan, Sztucki, Zinn, Kieffer, Homs-Puron, Gorini, Van~Vaerenbergh, and Boesecke]{narayanan2022}
Narayanan,~T.; Sztucki,~M.; Zinn,~T.; Kieffer,~J.; Homs-Puron,~A.; Gorini,~J.; Van~Vaerenbergh,~P.; Boesecke,~P. Performance of the time-resolved ultra-small-angle X-ray scattering beamline with the Extremely Brilliant Source. \emph{Applied Crystallography} \textbf{2022}, \emph{55}, 98--111\relax
\mciteBstWouldAddEndPuncttrue
\mciteSetBstMidEndSepPunct{\mcitedefaultmidpunct}
{\mcitedefaultendpunct}{\mcitedefaultseppunct}\relax
\EndOfBibitem
\bibitem[Sztucki(2021)]{sztucki2021SAXSutilities2}
Sztucki,~M. SAXSutilities2: a graphical user interface for processing and analysis of Small-Angle X-ray Scattering data. \emph{Zenodo} \textbf{2021}, \relax
\mciteBstWouldAddEndPunctfalse
\mciteSetBstMidEndSepPunct{\mcitedefaultmidpunct}
{}{\mcitedefaultseppunct}\relax
\EndOfBibitem
\bibitem[SasView()]{SasView}
SasView \url{http://www.sasview.org/}\relax
\mciteBstWouldAddEndPuncttrue
\mciteSetBstMidEndSepPunct{\mcitedefaultmidpunct}
{\mcitedefaultendpunct}{\mcitedefaultseppunct}\relax
\EndOfBibitem
\bibitem[Daoud and Cotton(1982)Daoud, and Cotton]{daoud1982star}
Daoud,~M.; Cotton,~J. Star shaped polymers: a model for the conformation and its concentration dependence. \emph{Journal de Physique} \textbf{1982}, \emph{43}, 531--538\relax
\mciteBstWouldAddEndPuncttrue
\mciteSetBstMidEndSepPunct{\mcitedefaultmidpunct}
{\mcitedefaultendpunct}{\mcitedefaultseppunct}\relax
\EndOfBibitem
\bibitem[Gnan \latin{et~al.}(2017)Gnan, Rovigatti, Bergman, and Zaccarelli]{gnan2017silico}
Gnan,~N.; Rovigatti,~L.; Bergman,~M.; Zaccarelli,~E. In silico synthesis of microgel particles. \emph{Macromolecules} \textbf{2017}, \emph{50}, 8777--8786\relax
\mciteBstWouldAddEndPuncttrue
\mciteSetBstMidEndSepPunct{\mcitedefaultmidpunct}
{\mcitedefaultendpunct}{\mcitedefaultseppunct}\relax
\EndOfBibitem
\bibitem[Rovigatti \latin{et~al.}(2015)Rovigatti, {\v{S}}ulc, Reguly, and Romano]{rovigatti2015comparison}
Rovigatti,~L.; {\v{S}}ulc,~P.; Reguly,~I.~Z.; Romano,~F. A comparison between parallelization approaches in molecular dynamics simulations on GPUs. \emph{Journal of computational chemistry} \textbf{2015}, \emph{36}, 1--8\relax
\mciteBstWouldAddEndPuncttrue
\mciteSetBstMidEndSepPunct{\mcitedefaultmidpunct}
{\mcitedefaultendpunct}{\mcitedefaultseppunct}\relax
\EndOfBibitem
\bibitem[Sciortino(2017)]{sciortino2017three}
Sciortino,~F. Three-body potential for simulating bond swaps in molecular dynamics. \emph{The European Physical Journal E} \textbf{2017}, \emph{40}, 1--4\relax
\mciteBstWouldAddEndPuncttrue
\mciteSetBstMidEndSepPunct{\mcitedefaultmidpunct}
{\mcitedefaultendpunct}{\mcitedefaultseppunct}\relax
\EndOfBibitem
\bibitem[Grest and Kremer(1986)Grest, and Kremer]{grest1986}
Grest,~G.~S.; Kremer,~K. Molecular Dynamics Simulation for Polymers in the Presence of a Heat Bath. \emph{Physical Review A} \textbf{1986}, \emph{33}, 3628\relax
\mciteBstWouldAddEndPuncttrue
\mciteSetBstMidEndSepPunct{\mcitedefaultmidpunct}
{\mcitedefaultendpunct}{\mcitedefaultseppunct}\relax
\EndOfBibitem
\bibitem[Soddemann \latin{et~al.}(2001)Soddemann, D{\"u}nweg, and Kremer]{soddemann2001}
Soddemann,~T.; D{\"u}nweg,~B.; Kremer,~K. A generic computer model for amphiphilic systems. \emph{The European Physical Journal E} \textbf{2001}, \emph{6}, 409--419\relax
\mciteBstWouldAddEndPuncttrue
\mciteSetBstMidEndSepPunct{\mcitedefaultmidpunct}
{\mcitedefaultendpunct}{\mcitedefaultseppunct}\relax
\EndOfBibitem
\bibitem[Thompson \latin{et~al.}(2022)Thompson, Aktulga, Berger, Bolintineanu, Brown, Crozier, In't~Veld, Kohlmeyer, Moore, Nguyen, \latin{et~al.} others]{thompson2022lammps}
Thompson,~A.~P.; Aktulga,~H.~M.; Berger,~R.; Bolintineanu,~D.~S.; Brown,~W.~M.; Crozier,~P.~S.; In't~Veld,~P.~J.; Kohlmeyer,~A.; Moore,~S.~G.; Nguyen,~T.~D.; others LAMMPS-a flexible simulation tool for particle-based materials modeling at the atomic, meso, and continuum scales. \emph{Computer physics communications} \textbf{2022}, \emph{271}, 108171\relax
\mciteBstWouldAddEndPuncttrue
\mciteSetBstMidEndSepPunct{\mcitedefaultmidpunct}
{\mcitedefaultendpunct}{\mcitedefaultseppunct}\relax
\EndOfBibitem
\bibitem[Mayer and Likos(2007)Mayer, and Likos]{mayer2007coarse}
Mayer,~C.; Likos,~C.~N. A Coarse-Grained Description of Star- Linear Polymer Mixtures. \emph{Macromolecules} \textbf{2007}, \emph{40}, 1196--1206\relax
\mciteBstWouldAddEndPuncttrue
\mciteSetBstMidEndSepPunct{\mcitedefaultmidpunct}
{\mcitedefaultendpunct}{\mcitedefaultseppunct}\relax
\EndOfBibitem
\bibitem[{Ruiz-Franco} \latin{et~al.}(2019){Ruiz-Franco}, {Jaramillo-Cano}, Camargo, Likos, and Zaccarelli]{ruiz-francoMultiparticleCollisionDynamics2019}
{Ruiz-Franco},~J.; {Jaramillo-Cano},~D.; Camargo,~M.; Likos,~C.~N.; Zaccarelli,~E. Multi-Particle Collision Dynamics for a Coarse-Grained Model of Soft Colloids. \emph{The Journal of Chemical Physics} \textbf{2019}, \emph{151}, 074902\relax
\mciteBstWouldAddEndPuncttrue
\mciteSetBstMidEndSepPunct{\mcitedefaultmidpunct}
{\mcitedefaultendpunct}{\mcitedefaultseppunct}\relax
\EndOfBibitem
\end{mcitethebibliography}
\newpage
\begin{suppinfo}
\setcounter{figure}{0}
\setcounter{table}{0}
\setcounter{page}{1}
\renewcommand{\theequation}{S\arabic{equation}}
\renewcommand{\thefigure}{S\arabic{figure}}
\renewcommand{\thetable}{S\arabic{table}}

\begin{center}
    \large\textbf{Star-like thermoresponsive microgels: a new class of soft nanocolloids} \\[1em]
    \normalsize Elisa Ballin$^{1,2,\dagger,*}$, Francesco Brasili$^{1,2,\dagger}$, Tommaso Papetti$^{1,2}$, Jacopo Vialetto$^{3,4}$, Michael Sztucki$^{5}$, Simona Sennato$^{1,2}$, Marco Laurati$^{3,4}$, Emanuela Zaccarelli$^{1,2,*}$ \\[0.5em]
    {\footnotesize $^1$ Dipartimento di Fisica, Sapienza Università di Roma, Piazzale A. Moro 2, 00185 Roma, Italy\\ $^2$ CNR-ISC, Uos Sapienza, Piazzale A. Moro 2, 00185 Roma, Italy \\ $^3$ Dipartimento di Chimica "Ugo Schiff", Università di Firenze,  Sesto Fiorentino (FI), 50019 Italy\\ $^4$ Consorzio per lo Sviluppo dei Sistemi a Grande Interfase (CSGI), via della Lastruccia 3, Sesto Fiorentino (FI), 50019, Italy\\ $^5$ European Synchrotron Radiation Facility – The European Synchrotron, 71 avenue des Martyrs F-38043 Grenoble, France \\ 
    $^{\dagger}$ These authors contributed equally\\
    $^*$Corresponding authors: elisa.ballin@uniroma1.it, marco.laurati@unifi.it, emanuela.zaccarelli@cnr.it}

\end{center}

\subsection{DLS characterisation}
In Figure ~\ref{fig:SI_RH} we report the hydrodynamic radius as a function of temperature for microgels synthesized with different contents of EGDMA from $0.5\%$ to $10\%$. We report also the curve for EGDMA $1\%$ synthesized with APS as initiator. As discussed in Results section, microgels retain a high swelling capacity and a very sharp transition also at high crosslinker content.
\begin{figure}[H]
    \centering
    \includegraphics[width=1\linewidth]{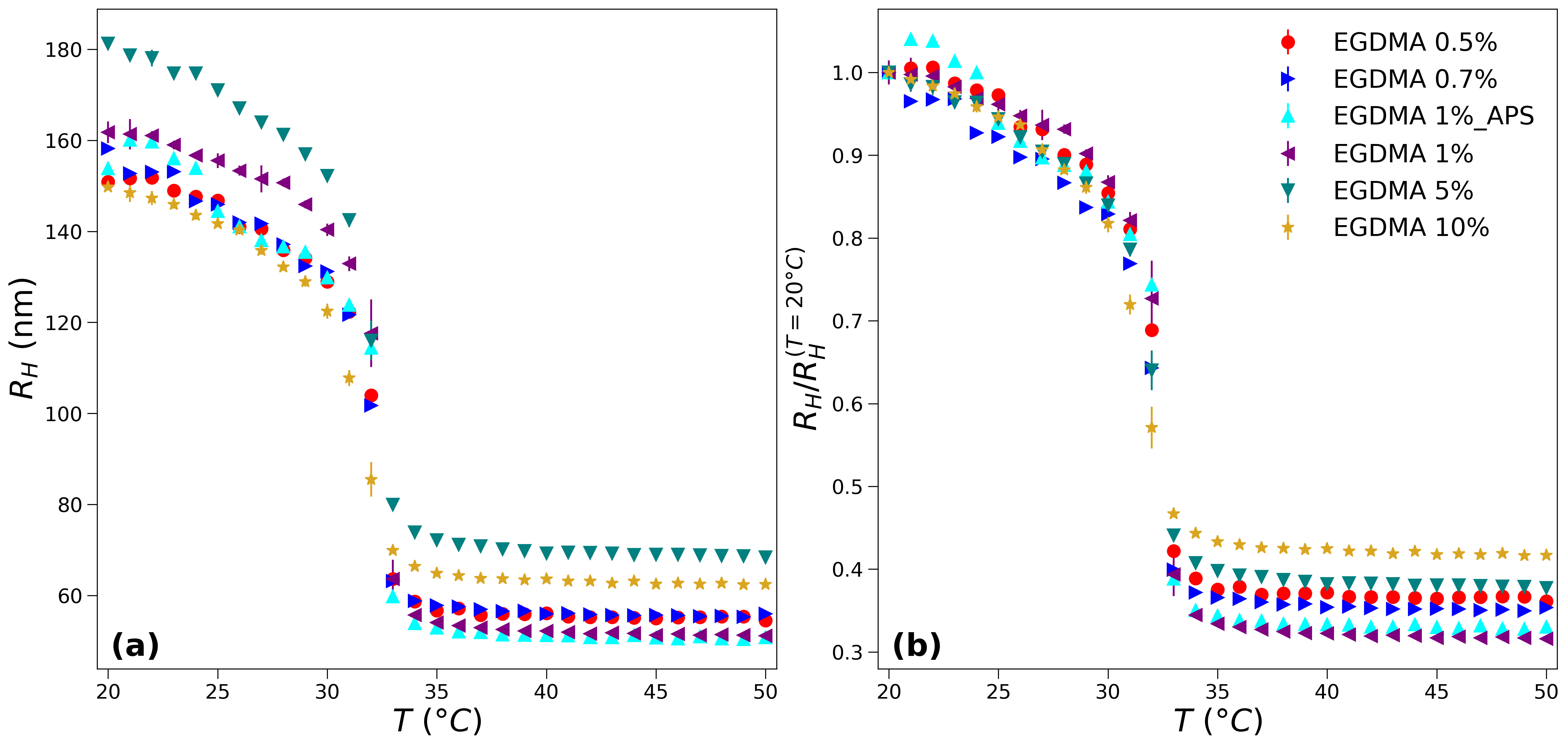}
    \caption{(a) Hydrodynamic radius as a function of temperature $T$ of PNIPAM-EGDMA microgels synthesized with different content of EGDMA and initiators; (b) corresponding normalized curves. }
    \label{fig:SI_RH}
\end{figure}

In Table ~\ref{tab:SI_DLS_values} we report the main characteristics of PNIPAM-EGDMA microgels as determined by DLS measurements.
\begin{table}[H]
    \centering
\begin{tabular}{|lccccc|}
\hline
         &&&&&  \\ 
         Sample & $R_H^{T=25\celsius}$ (nm) &  $R_H^{T=45\celsius}$ (nm) & $S_R$& $s$ $(\celsius^{-1})$& $T_c$ $(\celsius)$ \\
        &&&& & \\   \hline
         EGDMA $0.5\%$& $147 \pm 2 $&  $55.0\pm 0.2$  &$ 2.74 \pm 0.04$& $1.3\pm0.2$ & $32.17\pm0.07$ \\
         EGDMA $0.7\%$& $146 \pm 2 $ &  $55.7\pm0.2$  & $2.84\pm 0.03$ & $1.2\pm0.2 $& $32.09\pm0.07$ \\
         EGDMA $1\%$& $156 \pm 2$ &  $51.3 \pm 0.1$ & $3.15\pm0.05$& $1.5 \pm 0.2$& $32.28\pm0.08$ \\
         EGDMA $5\%$& $171\pm1$ & $ 68.9\pm0.3$ & $2.63\pm0.02$& $0.93\pm0.09$& $31.97\pm0.06$\\
         EGDMA $10\%$& $141.7\pm 0.5$ &  $62.5\pm0.1$ & $2.40\pm0.01$& $0.66\pm0.05$ &$31.41\pm0.03$ \\ \hline
    \end{tabular}
    \caption{Values of the hydrodynamic radius at $T=25\celsius$ and $T=45\celsius$, swelling ratio $S_R$, sharpness parameter $s$ and VPT temperature $T_c$ for microgels synthesized with $C_{EGDMA}$ from $0.5\%$, to $10\%$. }
    \label{tab:SI_DLS_values}
\end{table}

\subsection{Additional Results for PNIPAM-EGDMA microgels with $\mathbf{C_{EGDMA}=1\%}$}
In Fig.~\ref{fig:SI_EGDMA1_APS}(a) we show that the low-$T$ data for $C_{EGDMA}=1\%$ are also well-described by the simple star model of Dozier and coworkers~\cite{dozierColloidalNatureStar1991}, amounting to the sum of Eqns.~\ref{eq:star_model} and ~\ref{eq:guinier}.  We find $R_g\sim91$nm, $\mu\approx0.66$ and $\xi\sim20$nm and note that the parameters of the fit are largely identical to those reported in the main text (Table~\ref{tab:par_coreshellfuzzy}) when using the star-like fuzzy sphere model of Eq.~\ref{eq:model}. However, at high $T$, the star model is not sufficient to describe the experimental form factors, which is why we resort to the more general model in all cases.

In addition, we test the possible influence of a different initiator on the resulting synthesis. To this aim, we also synthesized PNIPAM-EGDMA microgels with $C_{EGDMA}=1\%$ using APS as initiator, using the same amounts of reagents reported in Table \ref{tab:reagents} and injecting 1.2 ml of a water solution containing 9.8 mg of APS to start the reaction. In Figure ~\ref{fig:SI_EGDMA1_APS} (b), we show the form factors of these microgels at $T=25\celsius$ and $T=45\celsius$. It can be seen that by varying the initiator, a similar structure to the KPS-initiated ones is obtained. At low temperature the microgels synthesized with APS display a small peak with respect to EGDMA $1\%$-KPS microgel. This small difference probably is due to a greater incorporation of EGDMA that may vary from one synthesis to another. At high temperature the form factors of the two microgels are perfectly superimposed.
The data at both temperatures are again well described by the star-like fuzzy sphere model of Eq.~\ref{eq:model}, with parameters reported in Table ~\ref{tab:SI_APS}, very similar to the ones corresponding to the synthesis in the presence of KPS reported in Table~\ref{tab:par_coreshellfuzzy} of the main text.
\begin{figure}[H]
    \centering
    \includegraphics[width=1\linewidth]{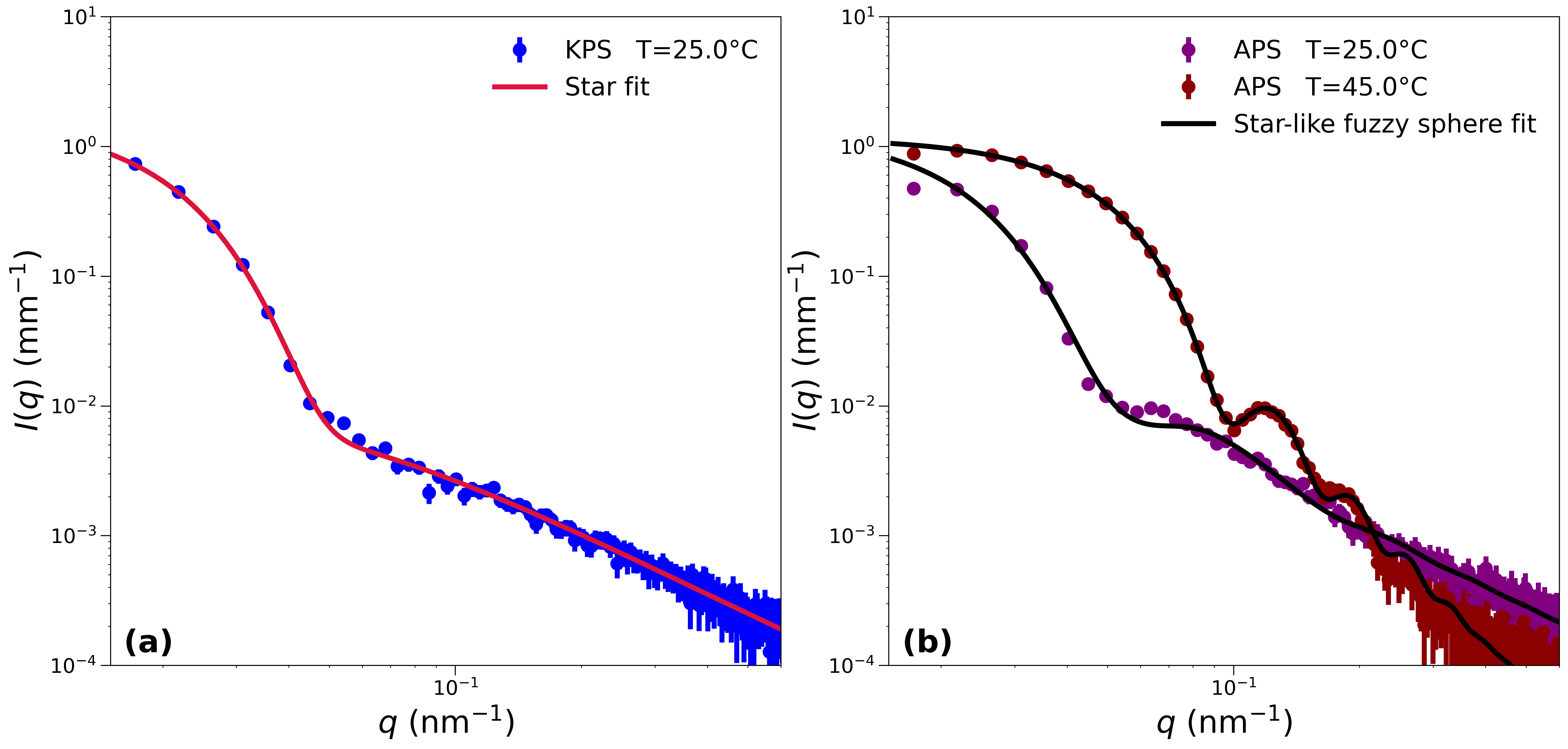}
    \caption{(a) SAXS scattering intensities I(Q) at $T=25\celsius$ for PNIPAM-EGDMA microgels with $C_{EGDMA}=1\%$ as in Fig.~\ref{fig:figura2}(b) of the main text, but now fitted with the simple star model proposed by Dozier \textit{et al.}~\cite{dozierColloidalNatureStar1991}; (b) I(Q) for $C_{EGDMA}=1\%$ microgels synthesized with APS as initiator at $T=25\celsius$ and $T=45\celsius$. The lines are fits via the star-like fuzzy sphere model in Eq.~\ref{eq:model}, whose fit parameters are reported in Table~\ref{tab:SI_APS}.}
    \label{fig:SI_EGDMA1_APS}
\end{figure}

\begin{table}[H]
    \centering
    \begin{tabular}{|cccccccc|}
    \hline
       T($\celsius$) & $R_c$(nm) & $t$(nm)  & $\sigma_s$(nm)& $\mu$ & $\xi$(nm)&$\Delta\rho^{cs}$(nm$^{-2}$)&$\Delta\rho^{s0}$(nm$^{-2}$)\\ \hline
      25 & 24 & 64 &28 & 0.66 & 20 & 0.1 &0.009 \\
      45 & 25 & 21 & 4 & 1.77 & 14  & 0.009 &0.04 \\
      \hline
    \end{tabular}
    \caption{Best-fit parameters for sample EGDMA $1\%$ synthesized with APS as initiator derived from the star-like fuzzy sphere model (Eq.~\ref{eq:model}). $\Delta\rho^{cs}$ and $\Delta\rho^{s0}$ represent the difference between the scattering length densities between the core and the shell ($\rho_c-\rho_s$) and between the shell and the solvent ($\rho_s-\rho_0$ ), respectively.}
    \label{tab:SI_APS}
\end{table}

\subsection{Star-polymer model for EGDMA $\mathbf{10\%}$}
In Figure \ref{fig:SI_EGDMA10_starpolymer} we show our attempts to model the form factors of EGDMA $10\%$ with an \textit{in silico} star polymer as the one used to describe EGDMA $1\%$. 
The \textit{in silico} star polymer  that has the most similar features to the experimental one was built using $R_c = 12.5 \sigma$, $f = 80$ and $N = 150$. However, as increasing the concentration of EGDMA leads to an increase in the size of the core, we cannot use a single large bead in the center, but we have to fill this randomly with a number of smaller  beads, as described in the Methods of the main text, to correctly capture the fact that experimentally, the core is full of crosslinkers as well as monomers. By adopting this modification, we have a somehow revisited star model, because (i) the core is quite large with respect to the total size of the particle and (ii) the core is not completely covered by arms on its surface. Notwithstanding this, 
the model has a similar shape and also the same peak positions as the experimental form factor at low temperature, as shown in Figure \ref{fig:SI_EGDMA10_starpolymer}(a). However, at high temperatures (b) the model completely fails, not being able to reproduce the main features of the experimental microgel.

\begin{figure}[H]
    \centering
    \includegraphics[width=1\linewidth]{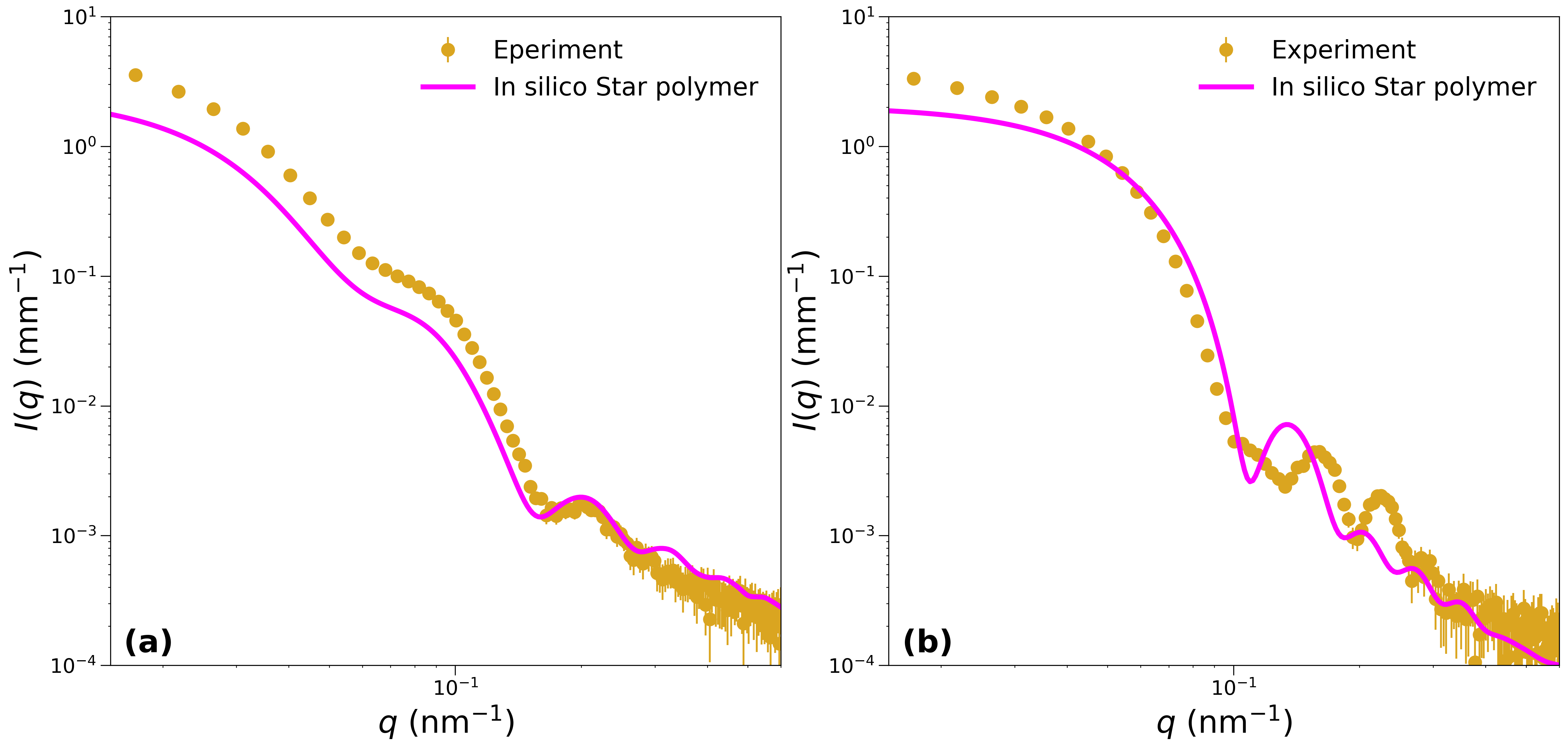}
    \caption{SAXS scattering intensities I(Q) at $T = 25\celsius$ (a) and $T=45\celsius$ (b) for samples with EGDMA $10\%$. The
solid lines represents the form factors of a simulated star-polymer at $\alpha=0.0$ (a) and $\alpha=0.8$ (b).}
    \label{fig:SI_EGDMA10_starpolymer}
\end{figure}

\subsection{Density profiles for \textit{in silico} star microgel and star polymer models}
In Figure \ref{fig:SI_densityprofile} we report the comparison between the density profiles of simulated EGDMA $1\%$ star-microgel and EGDMA $1\%$ star-polymer.
A clear difference is present in the region of the core, since in the case of the star polymer, there is a single rigid core onto which polymer chains are attached. In the case of the star microgel the core is full of crosslinkers. However, the tail region is similar in both models, following Eq.~\ref{eq:Rho_Star} as shown in the main manuscript, corroborating that the proposed method for modelling these microgels shows the features of a star polymer.
\begin{figure}[H]
    \centering
    \includegraphics[width=1\linewidth]{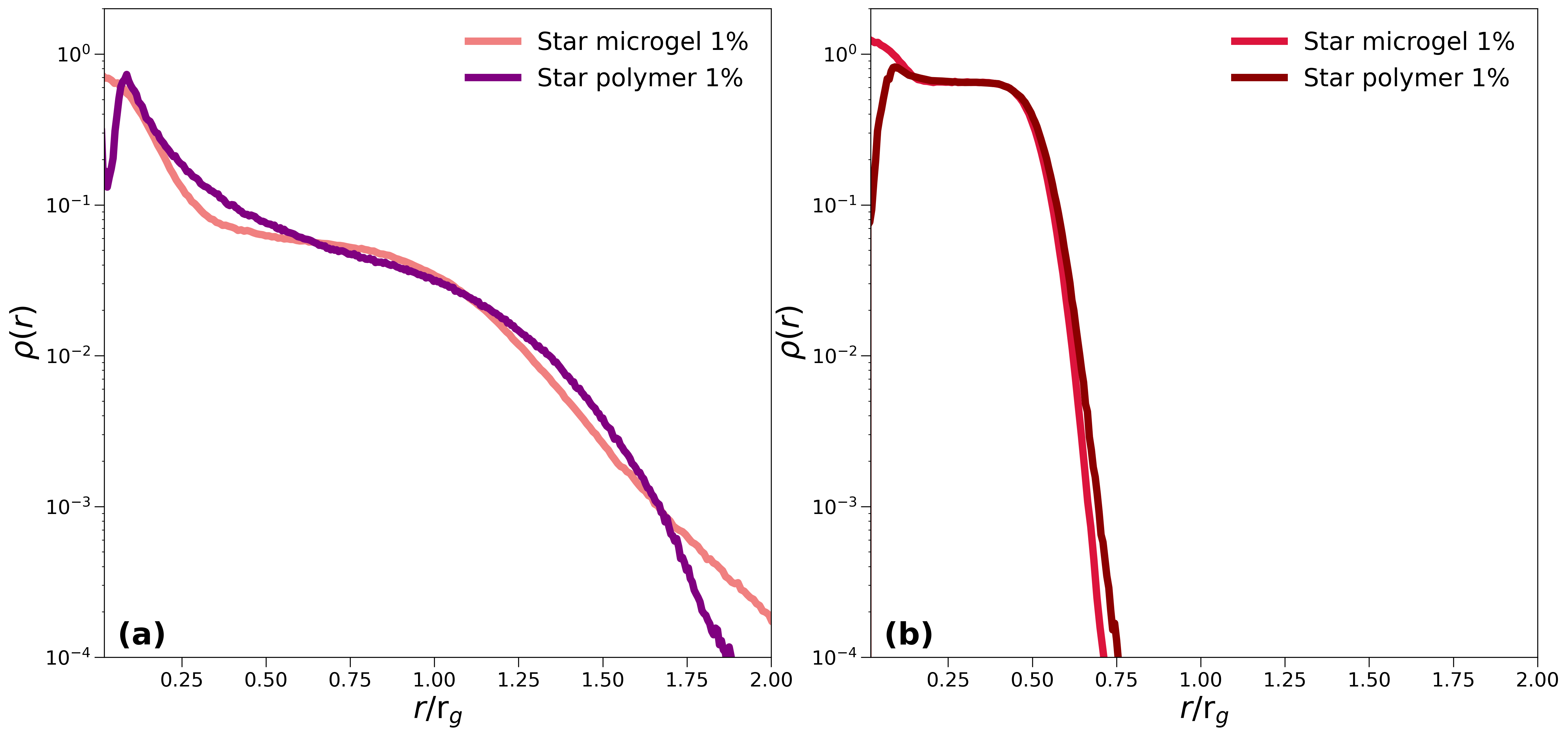}
    \caption{Comparison between the density profiles of simulated EGDMA $1\%$ star-microgel and EGDMA $1\%$ star-polymer at $\alpha=0.0$ (a) and $\alpha=0.8$ (b). The x-axis of each curve is rescaled by the radius of gyration calculated at $\alpha=0.0$.}
    \label{fig:SI_densityprofile}
\end{figure}

\subsection{\textit{In silico} swelling behaviour}
In Figure  \ref{fig:SI_Rg_alpha} we report the comparison between the swelling curves of simulated EGDMA $1\%$ and $10\%$ star-microgels and BIS $1\%$ and $10\%$ standard core-shell microgels. Unlike standard microgels, star-like ones retain a high swelling capacity even at high crosslinker content. Moreover the VPT remains very sharp if compared to the swelling curve of BIS $10\%$, even by not taking into account the different $\alpha$-temperature mapping that we established in the inset of Fig.~\ref{fig:figura6}(b).
\begin{figure}[H]
    \centering
    \includegraphics[width=0.5\linewidth]{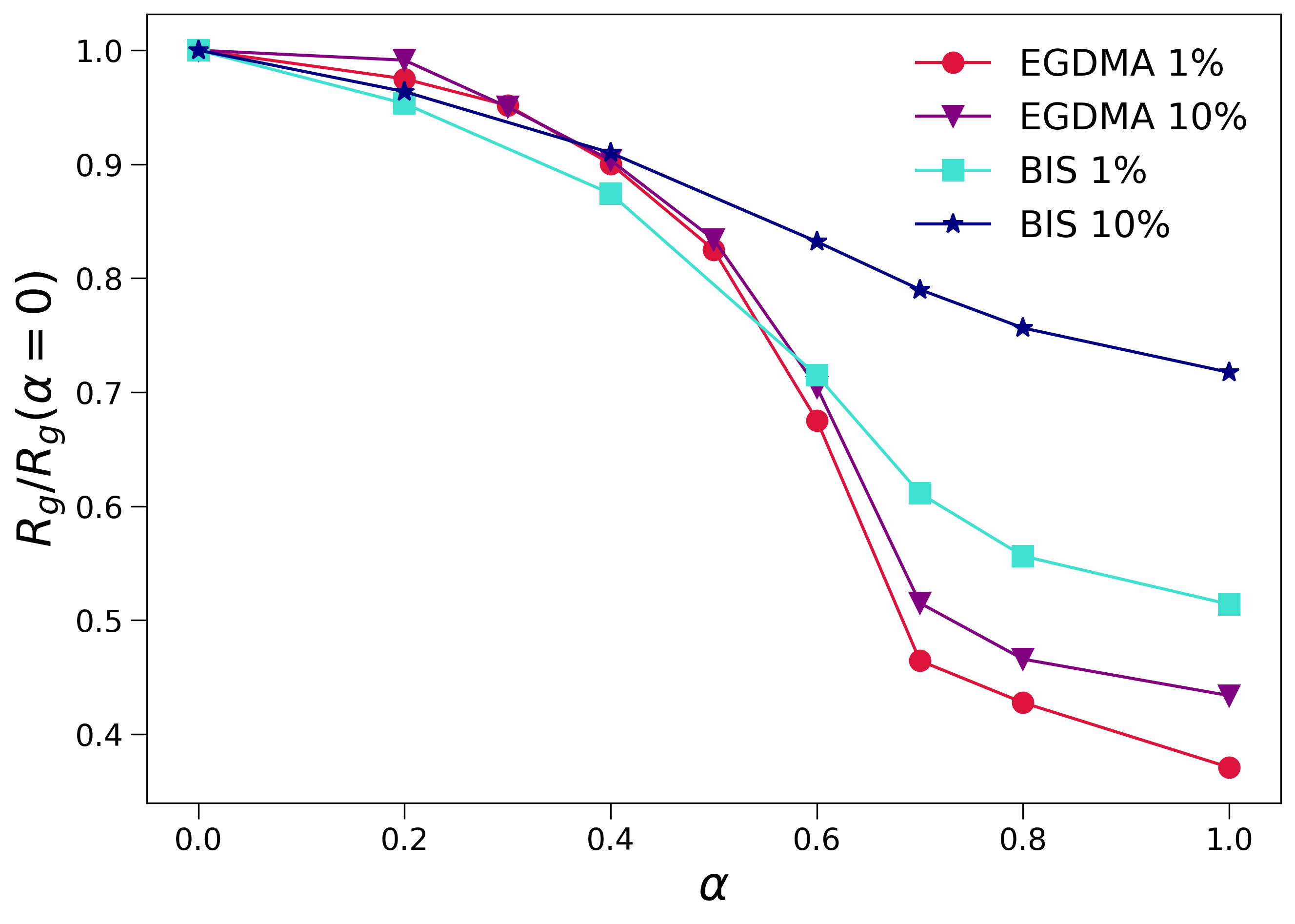}
    \caption{Comparison between swelling curves of simulated EGDMA $1\%$ and $10\%$ star-microgel and BIS $1\%$ and $10\%$ standard core-shell microgels. }
    \label{fig:SI_Rg_alpha}
\end{figure}

\end{suppinfo}

\end{document}